
\input epsf
\ifx\epsffile\undefined\message{(FIGURES WILL BE IGNORED)}
\def\insertfig#1#2{}
\else\message{(FIGURES WILL BE INCLUDED)}
\def\insertfig#1#2{{{\baselineskip=4pt
\midinsert\centerline{\epsfxsize=\hsize\epsffile{#2}}{{\centerline{#1}}}\
\medskip\endinsert}}}

\def\insertfigmed#1#2{{{\baselineskip=4pt
\midinsert\centerline{\epsfxsize=2.5in\epsffile{#2}}{{\centerline{#1}}}\
\medskip\endinsert}}}
\def\insertfigmedbig#1#2{{{\baselineskip=4pt
\midinsert\centerline{\epsfxsize=5.0in\epsffile{#2}}{{\centerline{#1}}}\
\medskip\endinsert}}}

\fi

\input harvmac
\input tables

%

%
\ifx\answ\bigans
\else
\output={
  \almostshipout{\leftline{\vbox{\pagebody\makefootline}}}\advancepageno }
\fi
%
%
%

%
%

%
%
\def\UCSD#1#2{\noindent#1\hfill #2%
\bigskip\supereject\global\hsize=\hsbody%
\footline={\hss\tenrm\folio\hss}}
%
%
\def\abstract#1{\centerline{\bf Abstract}\nobreak\medskip\nobreak\par #1}
%
%
%
%
\edef\tfontsize{ scaled\magstep3}
 \tfontsize  \tfontsize
 \tfontsize \font\titlei=cmmi10 \tfontsize
\font\titleis=cmmi7 \tfontsize \font\titleiss=cmmi5 \tfontsize
\font\titlesy=cmsy10 \tfontsize \font\titlesys=cmsy7 \tfontsize
\font\titlesyss=cmsy5 \tfontsize  \tfontsize
\skewchar\titlei='177 \skewchar\titleis='177 \skewchar\titleiss='177
\skewchar\titlesy='60 \skewchar\titlesys='60 \skewchar\titlesyss='60
%
%
%
%
%
\def\inv{^{\raise.15ex\hbox{${\scriptscriptstyle -}$}\kern-.05em 1}}
\def\lbar{{\lower.35ex\hbox{$\mathchar'26$}\mkern-10mu\lambda}} 
\def\e#1{{\rm e}^{^{\textstyle#1}}}

%
%
%
%
\def\dsl{\,\raise.15ex\hbox{/}\mkern-13.5mu D} 
\def\delsl{\raise.15ex\hbox{/}\kern-.57em\partial}
\def\Ksl{\hbox{/\kern-.6000em\rm K}}
\def\Asl{\hbox{/\kern-.6500em \rm A}}
\def\Dsl{\hbox{/\kern-.6000em\rm D}} 
\def\Qsl{\hbox{/\kern-.6000em\rm Q}}
\def\gradsl{\hbox{/\kern-.6500em$\nabla$}}
%
%
\def\lspace{\ifx\answ\bigans{}\else\qquad\fi}
\def\lbspace{\ifx\answ\bigans{}\else\hskip-.2in\fi} 
%
%
\def\boxeqn#1{\vcenter{\vbox{\hrule\hbox{\vrule\kern3pt\vbox{\kern3pt
        \hbox{${\displaystyle #1}$}\kern3pt}\kern3pt\vrule}\hrule}}}
%
%
\def\mbox#1#2{\vcenter{\hrule \hbox{\vrule height#2in
\kern#1in \vrule} \hrule}}
%
%
%
%

%
%
%
%
%

%

\def\bar#1{\overline{#1}}

\def\darr#1{\raise1.5ex\hbox{$\leftrightarrow$}\mkern-16.5mu #1}

%
%
\def\frac#1#2{{\textstyle{#1\over #2}}} 
%
%
%
%

%
%
%
%

%
%
\def\ltap{\ \raise.3ex\hbox{$<$\kern-.75em\lower1ex\hbox{$\sim$}}\ }
\def\gtap{\ \raise.3ex\hbox{$>$\kern-.75em\lower1ex\hbox{$\sim$}}\ }
\def\gl{\ \raise.5ex\hbox{$>$}\kern-.8em\lower.5ex\hbox{$<$}\ }
\def\roughly#1{\raise.3ex\hbox{$#1$\kern-.75em\lower1ex\hbox{$\sim$}}}
%
%

%

%
\def\np#1#2#3{{Nucl. Phys. } B{#1} (#2) #3}
\def\pl#1#2#3{{Phys. Lett. } {#1}B (#2) #3}
\def\prl#1#2#3{{Phys. Rev. Lett. } {#1} (#2) #3}
\def\physrev#1#2#3{{Phys. Rev. } {#1} (#2) #3}

\relax

\def\lta{\ \hbox{\raise.55ex\hbox{$<$}} \!\!\!\!\!
\hbox{\raise-.5ex\hbox{$\sim$}}\ }
\def\gta{\ \hbox{\raise.55ex\hbox{$>$}} \!\!\!\!\!
\hbox{\raise-.5ex\hbox{$\sim$}}\ }

\def\qsl{\hbox{/\kern-.5600em {$q$}}}
\def\ksl{\hbox{/\kern-.5600em {$k$}}}

\def\({\left(}
\def\){\right)}

\def\L{{\cal L}}

\def\OMIT#1{}
\def\frac#1#2{{#1\over#2}}

\def\K{K_L \rightarrow \pi^+ \pi^- e^+ e^-}
\def\Kgam{K_L~\rightarrow~\pi^+~\pi^-~\gamma^*}
\def\e{\rightarrow \pi^+ \pi^- e^+ e^-}
\def\L{{\cal L}}
\def\k{\hbox{/\kern-.5600 em {$k$}}}
\def\PP{(\vec p_- \times \vec p_+)
    \cdot (\vec k_- - \vec k_+)}

\hbadness=10000

\noblackbox
\vskip 1.in
\centerline{{\titlefont{$K_L \rightarrow \pi^+ \pi^- e^+ e^-$}
\footnote{*}{{\tenrm Work
supported in part by the Department of Energy under contracts
DE--FG02--91ER40682 (CMU) and DE-FG03-92-ER40701 (Caltech) .}}}}
\vskip .5in
\centerline{John K. Elwood and Mark B. Wise}
\medskip
\centerline{\it California Institute of Technology, Pasadena, CA 91125}
\medskip
\vskip .25in
\centerline{\it and }
\medskip
\vskip .25in
\centerline{Martin J. Savage}
\medskip
\centerline{\it Department of Physics, Carnegie Mellon University,
Pittsburgh PA 15213}

\vskip .2in

\abstract{
We calculate all of the form factors for the one-photon,
$K_L \rightarrow \pi^+ \pi^- \gamma^* \rightarrow \pi^+ \pi^- e^+e^-$
contribution to the $K_L \rightarrow \pi^+ \pi^- e^+e^-$ decay amplitude
at leading  order in chiral perturbation theory.
These form factors depend on one unknown constant that is a linear
combination of coefficients of local ${\cal O}(p^4)$ operators in the
chiral lagrangian for weak radiative kaon decay.
We determine the differential rate for $\K$ and also the
magnitude of  two CP violating observables.
 }

\vfill
\UCSD{\vbox{
\hbox{CALT-68-1980}
\hbox{CMU-HEP 95-02}
\hbox{DOE-ER/40682-91}
\hbox{DE-FG03-92-ER40701}}
}{April 1995}
\eject

\newsec{Introduction}

At the present time about twenty $\K$ events have been observed
and a detailed experimental study of this decay mode will be possible in future
experiments
\ref\wah{Y.W. Wah, in {\it Proceedings of the XXVI International Conference on
High
Energy Physics}, Dallas, Texas, AIP Conf. Proc. No. 272, edited by James R.
Sanford
(AIP, New York) , 1992.}.
The $\K$ weak decay amplitude is dominated by the process,
$\Kgam \e $, where a single virtual photon creates the $e^+e^-$ pair.
This one photon contribution to the decay  amplitude has the form
\eqn\onegam{
M^{(1\gamma)} = { s_1 G_F  \alpha \over 4 \pi  f  q^2}
\left[  i G \varepsilon^{\mu\lambda \rho \sigma} p_{+ \lambda}\  p_{- \rho}\
q_\sigma \ +\  F_+ p_+^\mu \ +\  F_- p_-^\mu\ \right]
\cdot \bar u (k_-) \gamma_\mu v(k_+)\ \ \ \ \ ,}
where $G_F$ is Fermi's constant, $\alpha$ is the electromagnetic fine
structure constant,  $s_1 \simeq 0.22$ is the sine of the Cabibbo angle
and $f
\simeq 132$ MeV is the pion decay constant.  The $\pi^+$ and $\pi^-$
four-momenta are denoted by $p_+$ and $p_-$ and the $e^+$ and $e^-$
four-momenta are denoted by $k_+$ and $k_-$.  The sum of the electron
and positron four-momenta is $q = k_- + k_+$.
The Lorentz scalar form factors $G, F_\pm$
depend on the scalar products of the four-momenta  $q , p_+$ and $p_-$.
Neglecting CP nonconservation, under interchange of the pion
four-momenta
\eqn\cpmom{
p_+ \rightarrow p_-\ \ {\rm and}\ \ \  p_- \rightarrow p_+ }
the form factors become
\eqn\cpform{
G \rightarrow G\ \ , \qquad F_+ \rightarrow F_-\ \ , \qquad F_- \rightarrow
F_+\ \ \ \ .}
In this paper we compute the CP conserving contribution to the form
factors $G, F_\pm$ using chiral perturbation theory at one-loop order
(the ${\cal O}(p^2)$ amplitude vanishes).
The coefficients of some of the local operators appearing at the same
order in the chiral expansion
(i.e., order $p^4$ counter terms, where $p$ is a typical momentum)
are determined by the experimental value of the pion
charge radius and the measured
$K^+ \rightarrow \pi^+ e^+ e^-$ and
$K_L \rightarrow \pi^+ \pi^- \gamma$ decay rates and spectra.

We also compute (in chiral perturbation theory) an important tree level
contribution to the
form factors $F_\pm$  that arises from the small CP even component of
the $K_L$ state.
This contribution to the $F_\pm$ form factors  from indirect CP
nonconservation  has the
opposite symmetry property under interchange of pion momenta
when compared with  the CP conserving contribution to $F_\pm$
(see eqs. $\cpmom$ and $\cpform$) .
If
\eqn\pimoma{
p_+ \rightarrow p_-\ \ \ {\rm and} \ \ \  p_- \rightarrow p_+ }
then the CP violating one-photon form factors become
\eqn\fsyma{
F_+ \rightarrow -F_-\ \ \ ,\ \ \ \  F_- \rightarrow - F_+\ \ \ .}

The decay amplitude that follows from squaring the invariant matrix
element in eq. (1.1) and summing over $e^+$ and $e^-$ spins is
symmetric under
interchange of  $e^+$ and $e^-$ momenta, $k_-\leftrightarrow k_+$.
Physical variables that are
antisymmetric under interchange of the  $e^+$ and $e^-$ momenta
arise from the interference of the short distance  contributions
($Z$-penguin and $W$-box diagrams) and the two photon piece with the
one photon amplitude given in eq. (1.1).

In the minimal standard model the coupling of the quarks to the
$W$-bosons has
the form
\eqn\smlag{
\L_{int} = - {g_2\over\sqrt{2}}\  \bar u_L^j \gamma_\mu V^{jk} d_L^k\
W^\mu
\ +\  h.c.  \ \ \ \ .}
Here repeated generation indices $j,k$ are summed over 1,2,3 and
$g_2$ is the
weak SU(2) gauge coupling.  $V$ is a $3\times 3$ unitary matrix (the
Cabibbo--Kobayashi--Maskawa matrix) that arises from diagonalization
of the
quark mass matrices.  By redefining phases of the quark fields it is
possible
to write $V$ in terms of four angles $\theta_1, \theta_2, \theta_3$
and $\delta$.  The $\theta_j$ are analogous to the Euler angles and
$\delta$ is a
phase that, in the minimal standard model, is responsible for the
observed CP violation.  Explicitly
\eqn\ckmmat{
V = \left(  \matrix{c_1 &-s_1 c_3 & -s_1s_3\cr
s_1 c_2 & c_1 c_2 c_3 - s_2 s_3 e^{i\delta} & c_1 c_2 s_3 + s_2 c_3
e^{i\delta}\cr
s_1 s_2 & c_1 s_2 c_3 + c_2 s_3 e^{i\delta} & c_1 s_2 s_3 - c_2 c_3
e^{i\delta}\cr}\right) \ \ \ ,}
where $c_i \equiv \cos\theta_i$ and $ s_i \equiv \sin\theta_i$.  It is
possible
to choose the $\theta_j$ to lie in the first quadrant.  Then the quadrant
of
$\delta$ has physical significance and cannot be chosen by a phase
convention
for the quark fields.  A value of $\delta$ not equal to $0$ or $\pi$
gives rise to CP violation.

The short distance $W$-box and $Z$-penguin Feynman diagrams depend
on the
$V_{ts}$ element  of the Cabibbo--Kobayashi--Maskawa matrix.  It is very
important to be able to determine this coupling experimentally.  In this paper
we calculate the contribution to the $K_L \e$ decay amplitude arising from
the $Z$-penguin and $W$-box diagrams which can be determined using chiral
perturbation
theory since the left handed current $\bar s \gamma_\mu (1 - \gamma_5) d$
is related to a generator of chiral symmetry.  At the present time all observed
CP nonconservation has its origin in $K^0 - \bar K^0$ mass mixing.
A  CP violating variable can be constructed in the decay $\K$ that gets an
important contribution from CP nonconservation in the $Z$-penguin and
$W$-box diagrams, that is, direct CP violation.
The variable (in the $K_L$ rest frame)
\eqn\cpvar{
A_{CP} = < { \PP \over | \PP |} >\ \ ,}
is even under charge conjugation and odd under parity.  It is also odd under
interchange of $\vec k_+$ and $\vec k_-$.  The real and imaginary
parts of $V_{ts}$ are comparable, and hence the CP conserving and CP
violating
parts of the $Z$-penguin and $W$-box diagrams are of roughly equal
importance.
 $A_{CP}$ gets a significant contribution from this direct source of CP
nonconservation.  In this  paper we calculate $A_{CP}$ in the minimal standard
model but unfortunately we find that it is quite  small; $|A_{CP}| \approx
10^{-4}$.

The decay $\K$ has been  studied previously by Sehgal and Wanninger
\ref\sehwan{L.M. Sehgal and M. Wanninger, \physrev{D46}{1992}{1035}.}
and  by Heiliger and Sehgal
\ref\heilseh{P. Heiliger and L.M. Sehgal, \physrev{D48}{1993}{4146}.}.
These authors adopted a phenomenological approach,
relating the $\K $ decay amplitude to the measured $K_L \rightarrow \pi^+
\pi^- \gamma$ decay amplitude.
In the systematic expansion of chiral perturbation theory we
find important additional contributions to the $\K$ decay amplitude
for $q^2~=~(k_-~+~k_+)^2~>>~4m_e^2$ that
were not included in this previous work. It
was pointed out in refs. \sehwan\   and \heilseh\  that indirect CP
nonconservation from $K^0-\bar K^0$ mixing gives an important contribution
to the $K_L \rightarrow \pi^+ \pi^-e^+e^-$ decay rate and consequently
there is a CP violating observable, $B_{CP}$,  that is quite large. We
reexamine $B_{CP}$ using the form factors determined in this paper.

\newsec{The One-Photon Amplitude}

Chiral perturbation theory provides a systematic approach to understanding
the
one-photon part of the $K_L \e$ decay amplitude.  It uses an effective
field theory that incorporates  the SU(3)$_L \times$SU(3)$_R$ chiral
symmetry of  QCD and an
expansion in powers of momentum to reduce the number of operators that
occur.
In the chiral Lagrangian the $\pi$'s , $K$'s and $\eta$ are incorporated
into a  $3\times 3$ special unitary matrix
\eqn\sigmes{
\Sigma = \exp \left({2iM\over f}\right)\ \ \ \ ,}
where
\eqn\mesons{
M = \left( \matrix{\pi^0/\sqrt{2} + \eta/\sqrt{6} & \pi^+ & K^+\cr
\pi^- & - \pi^0/\sqrt{2} + \eta/\sqrt{6} & K^0\cr
K^- & \bar K^0 & -2 \eta/\sqrt{6}\cr}\right) \ \ \ .}
At leading order in chiral perturbation theory $f  \simeq 132$ MeV is the pion
decay constant.
Under SU(3)$_L \times$ SU(3)$_R$ transformations the $\Sigma$ field
transforms as
\eqn\sigtrans{
\Sigma \rightarrow L \Sigma R^\dagger\ \ \ \ ,}
where
$L\ \epsilon\ $ $SU(3)_L$ and $R\ \epsilon\ $ $SU(3)_R$.

At leading order in chiral perturbation theory (i.e., order $p^2$, where $p$ is
a typical four-momentum) the strong and electromagnetic interactions of the
pseudo--Goldstone bosons are described by the chiral Lagrange density
\eqn\lagpi{
\L_S^{(1)} = {f^2\over 8} Tr (D_\mu \Sigma D^\mu \Sigma^\dagger) + v Tr (m_q
\Sigma + m_q \Sigma^\dagger)\ \ \ ,}
where $v$ is a parameter with dimensions of mass to the third power and $m_q$
is the quark mass matrix
\eqn\massmes{
m_q = \left( \matrix{m_u & 0 & 0\cr 0 & m_d & 0\cr 0 & 0 & m_s\cr}\right) \ \ \
.}
In this paper we neglect isospin violation in the quark mass matrix and set
$m_u = m_d$.  In this approximation the $K^0$ and $K^+$ have equal masses
which we denote by $m_K$, and the Gell-Mann--Okubo mass relation
\eqn\gellmann{
3 m_\eta^2 - 4  m_K^2 + m_\pi^2 = 0\ \ \ \ ,}
holds.

The effective Lagrangian for $\Delta S = 1$ weak nonleptonic decays
transforms
as $(8_L, 1_R) + (27_L, 1_R)$ under SU(3)$_L\otimes $SU(3)$_R$.
The $(8_L, 1_R)$ amplitudes are much larger  than the $(27_L, 1_R)$
amplitudes
and so  we will neglect the $(27_L, 1_R)$ part of the effective Lagrangian.
The effective Lagrangian for weak radiative kaon decay is obtained by gauging
the  effective
Lagrangian for weak nonleptonic decays with respect to the U(1)$_Q$ of
electromagnetism.
At leading order in  chiral perturbation theory the $\Delta S = 1$ transitions
are
described by
\eqn\delSone{
\L_W^{(1)} = {g_8 G_F s_1f^4\over 4\sqrt{2}}
Tr \left[  D_\mu\  \Sigma D^\mu \ \Sigma^\dagger \  T \right]\ +\ h.c.\ \ \ \ \
\ .}
The matrix $T$ in \delSone\  projects out the correct flavour structure of the
octet
\eqn\flavproj{
T = \left( \matrix{0 & 0 & 0\cr 0 & 0 & 0 \cr 0 & 1 & 0\cr}\right)\ \ \ \ ,}
and $g_8$ is a constant determined by the measured
$K_S \rightarrow \pi^+ \pi^-$  decay rate;  $|g_8| \simeq 5.1$.
In \lagpi\  and \delSone\  $D_\mu$ represents a covariant derivative:
\eqn\covder{
D_\mu \Sigma = \partial_\mu \Sigma + i e A_\mu [Q, \Sigma]\ \ \ ,}
where
\eqn\charge{
Q = \left( \matrix{2/3 & 0 & 0\cr 0 & -1/3 & 0\cr 0 & 0 & -1/3\cr}\right) \ \ \
\ ,}
is the electromagnetic charge matrix for the three lightest quarks, $u , d$ and
$s$.

The $K_L$ state
\eqn\klong{
|K_L> \simeq |K_2> + \epsilon |K_1>\ \ \ ,}
is mostly the CP odd state
\eqn\kcpodd{
|K_2> = {1\over\sqrt{2}} (|K^0> + |\bar K^0>)\ \ \ ,}
with small admixture of the CP even state
\eqn\kcpeven{
|K_1> = {1\over\sqrt{2}} (|K^0> - |\bar K^0>)\ \ \ .}
The parameter $\epsilon$ characterizes CP nonconservation in $K^0 - \bar
K^0$ mixing.  At leading order in chiral perturbation theory the
$\Kgam\e$  decay
amplitude arises though the CP even component of $K_L$.  Writing the form
factors  contributing to
$\Kgam$ as a power series in the chiral expansion
\eqn\powform{
F_\pm = F_\pm^{(1)} + F_\pm^{(2)}+...\ \  , \quad G = G^{(1)} + G^{(2)}
+...\ \ \  , }
where the superscript denotes the order of chiral perturbation theory, we
find that the Feynman diagrams in
\fig\epkppgam{Feynman diagrams contributing to $F_\pm^{(1)}$.}
give
\eqn\formep{\eqalign{
G^{(1)} & = 0 \cr
F_+^{(1)} &= - {32g_8f^2 (m_K^2 - m_\pi^2) \pi^2 \epsilon\over
[q^2 + 2q \cdot p_+]}\cr
F_-^{(1)} &= + {32g_8f^2 (m_K^2 - m_\pi^2) \pi^2 \epsilon\over [q^2 + 2 q
\cdot p_-]}
}\ \ \ \ .}
Despite the fact that $\epsilon \simeq 0.0023\ e^{i44^{o}}$ (in a phase
convention  where
the $K^0 \rightarrow \pi \pi (I = 0)$ decay amplitude is real), is small it is
important to keep this part of the decay amplitude.  Other contributions not
proportional to $\epsilon$ don't occur until higher order in chiral
perturbation theory.
We neglect direct sources of CP nonconservation in the one-photon part
of the decay amplitude.  Experimental information on $\epsilon^\prime$
suggests that they are small.

At the next order in the chiral expansion the form factors $G^{(2)},
F_\pm^{(2)}$ arise from  ${\cal O} ( p^4) $ local operators and from
one-loop  Feynman
diagrams involving vertices from the leading Lagrange densities in \lagpi\
and \delSone\ .
However, the form factor $G^{(2)}$ arises solely from local operators as the
one loop Feynman diagrams and tree graphs involving the
Wess--Zumino term
\ref\enp{G. Ecker, H. Neufeld and A. Pich, \pl{278}{1992}{337}.}
\ref\bea{J. Bijnens, G. Ecker and A. Pich, \pl{286}{1992}{341}.}
do not contribute.
The contribution of the
${\cal O} ( p^4) $ local operators to $G^{(2)}$ is fixed by the measured $K_L
\rightarrow \pi^+ \pi^- \gamma$ decay rate
\ref\donald{G. Donaldson {\it et al}  , \prl{33}{1974}{554};
\physrev{D14}{1976}{2839}.}
\ref\ramberg{E.J. Ramberg {\it et al} , \prl{70}{1993}{2525}.}
 to be
\eqn\gfix{
|G^{(2)}| \simeq 40\ \ \ .}
The experimentally observed $K_L \rightarrow \pi^+ \pi^- \gamma$
Dalitz plot suggests that the form
factor $G$ has significant momentum dependence.  This indicates that
$G^{(3)}$
is not negligible, and that our extraction of $G^{(2)}$ from the rate is not
completely justified
\ref\enpenp{G. Ecker, H. Neufeld and A. Pich, \np{413}{1994}{321}.}.

The form-factors $F_\pm^{(2)}$ get contributions both from local operators
of ${\cal O} ( p^4) $
\ref\ekw{G. Ecker, J. Kambor and D. Wyler, \np{394}{1993}{101}.}
and from one-loop diagrams involving vertices from the
leading
Lagrange densities in \lagpi\  and \delSone\ .  For $\K$ the local operators
that  contribute are
\eqn\picounter{
\L_S^{(2)} = {-ie \lambda_{cr} (\mu)\over 16\pi^2} F^{\mu\nu} Tr \left[ Q
(D_\mu
\Sigma
D_\nu \Sigma^\dagger + D_\mu \Sigma^\dagger D_\nu \Sigma)\right]  \ \ \ ,}
and
\eqn\kpipicounter{\eqalign{
\L_W^{(2)} = i\ {G_F s_1 e f^2 g_8\over \sqrt{2}~ 16 \pi^2}  &
\left[ \ \
a_1 (\mu) F^{\mu\nu} Tr [QT (\Sigma D_\mu \Sigma^\dagger)
(\Sigma D_\nu \Sigma^\dagger) ] \right.\cr
&\left. \ \  + a_2 (\mu) F^{\mu\nu}
Tr [Q (\Sigma D_\mu \Sigma^\dagger) T (\Sigma
D_\nu \Sigma^\dagger)]  \right. \cr
&\left. \ \ + a_3(\mu) F^{\mu\nu} Tr [ T [Q,\Sigma] D_\mu\Sigma^\dagger
\Sigma D_\nu\Sigma^\dagger
- T D_\mu\Sigma D_\nu\Sigma^\dagger\Sigma
[\Sigma^\dagger , Q] ] \right.\cr
&\left. \ \ + a_4(\mu)  F^{\mu\nu} Tr [ T\Sigma D_\mu\Sigma^\dagger [Q, \Sigma]
D_\nu\Sigma^\dagger ]
\ \ \right]  + h.c.}
\ \ \ .}
The coefficients $\lambda_{cr}, a_1 , a_2 , a_3$ and $a_4$ depend on the
renormalization
procedure used and we employ dimensional regularization with $\overline{MS}$
subtraction.  The dependence of the coefficients $\lambda_{cr}, a_{1,2,3,4}$ on
the
subtraction point $\mu$ cancels that coming from the one-loop diagrams.
Note that the basis of operators in eq. \kpipicounter\  is slightly different
than that used in
\ekw\ .  With this basis of operators the combination of counterterms
\eqn\wldef{w_L = a_3-a_4\ \ \ ,}
 is  independent of the subtraction point $\mu$ at one loop.

The value of $\lambda_{cr}$ is fixed by the measured $\pi^+$ charge radius;
$<r^2_\pi > = 0.44 \pm 0.02 ~{\rm fm}^2$.
The one-loop diagrams in
\fig\pioncharge{Feynman diagrams contributing to the $\pi^\pm$ charge radius,
$<r^2_\pi >$ ,  at leading order in chiral perturbation theory.}
give (using  $\overline{MS}$)
\eqn\fixrpi{
\lambda_{cr} (\mu) = - \left({2\pi^2\over 3} \right) f^2 <r_\pi^2>\ \  -\ \
{1\over
24} \left[ 2\  \ell n (m_\pi^2/\mu^2) +  \ell n (m_K^2/\mu^2)  \right]\ \  \ ,}
which implies that (at the subtraction point
$\mu =  1$GeV)
\eqn\lamfix{
\lambda_{cr} (1 GeV) = - 0.91\pm 0.06\ \ \ \ \ .}
A linear combination of $a_1$ and $a_2$ is fixed by the measured $K^+
\rightarrow \pi^+ e^+ e^- $ decay amplitude.  Fortunately it is the same
combination of $a_1$ and $a_2$ that enters into the $K_L \rightarrow \pi^+
\pi^- e^+ e^-$ decay amplitude.  The one-photon part of the $K^+ \rightarrow
\pi^+ e^+ e^-$ decay amplitude can be written in terms of a single form factor
$f (q^2)$
\eqn\kpgamp{
M^{(1\gamma)} (K^+ \rightarrow \pi^+ e^+ e^-) = {s_1 G_F\over\sqrt{2}} ~
{\alpha\over 4\pi} ~ f (q^2)\  p_\pi^\mu\  \bar u (k_-) \gamma_\mu v (k_+)\ \ \
\ .}
The one-loop diagrams in
\fig\kplus{The Feynman diagrams contributing to the amplitude for
$K^+\rightarrow\pi^+\gamma^*$ at leading order in chiral perturbation theory.
The solid square denotes a vertex from the gauged weak lagrangian in \delSone\
,
the solid circle denotes a vertex from the gauged strong lagrangian in \lagpi\
{}.
Figures 3(a) involve only weak and electromagnetic vertices while Figures 3(b)
also has a strong vertex.   Figures 3(c) are the contributions from the kaon
and
pion charge radii (including both loop graphs and the tree-level counterterm).
Figure 3(d) is the contribution of the weak counterterm as given by
\kpipicounter\ .
We have not shown the wavefunction renormalization of the tree graphs for the
process as the sum of these graphs vanish.}
and the operators in \picounter\  and \kpipicounter\
give
\ref\epr{G. Ecker, A. Pich and E. deRafael, \np{291}{1987}{692}.}
\eqn\kplusform{\eqalign{
f (q^2) = 2 g_8 & \left(   \phi_K (q^2) + \phi_\pi (q^2)  - {1\over 6}
\ell n (m_K^2/\mu^2) - {1\over 6} \ell n (m_\pi^2/\mu^2) \right. \cr
& \left. \ \  + \  {2\over 3} (a_1(\mu) + 2a_2(\mu))  -  4 \lambda_{cr}(\mu)
+ {1\over 3} \right) \cr
= 2 g_8 & \left( \   \phi_K (q^2) + \phi_\pi (q^2)  + w_+ \  \right)
} \ \ \ ,}
where
\eqn\kplusfunc{
\phi_i (q^2) = \int_0^1 dx \left({m^2_i\over q^2} - x (1 -
x)\right) \ell n \left( 1 - {q^2\over m^2_i} x (1 - x)\right)\ \ \ \ .}
This relation defines the  $\mu$ independent constant $w_+$
\epr\ which has been experimentally determined to be
\ref\zeller{C. Alliegro etal, \prl{68}{1992}{278}.}
\eqn\wplusnum{
w_+ = 0.89^{+0.24}_{-0.14}\ \ \ .}
Using the central values of $\lambda_{cr}(1 GeV)$ and $w_+$
we find that
\eqn\kpipifix{
a_1 (1GeV) + 2a_2 (1GeV) = -  6.0  \ \ \ ,}
with an associated error around $10\%$ (which is correlated with the
uncertainty in $\lambda_{cr}(1 GeV)$).
Throughout the remainder of this work we will use the central values of
$\lambda_{cr}(1 GeV)$ and $a_1 (1GeV) + 2a_2 (1GeV)$ and suppress the
associated  uncertainties.
Note that the contributions $\lambda_{cr}(1GeV)$ and
$(a_1 + 2a_2)(1GeV)$ to $f(q^2)$
are separately quite large but they almost cancel against each other.

At ${\cal O}(p^4)$  the form factors $F_\pm^{(2)}$ for $\K$ decay
follow from the Feynman diagrams in
\fig\kpipigam{Feynman diagrams contributing to the CP conserving
amplitude for  $\Kgam$ at
leading  order in  chiral perturbation theory.  The notation is the same as in
\kplus\
and we have not shown the wavefunction renormalization of the tree graphs for
the
process as the sum of these graphs vanish.}
and tree level matrix elements of the operators in \picounter\  and
\kpipicounter\ .
We find, using  $\overline{MS}$ subtraction, that
\eqn\bigbas{\eqalign{
F_-^{(2)} = g_8 & \left(
 - {2\over 3} q^2 [a_1 (\mu) + 2a_2 (\mu) + 6 a_3(\mu) - 6 a_4(\mu)  ]
- 4 q^2 \lambda_{cr}(\mu)
+ {2\over 3} q^2 +  \phi_{K\eta} + \phi_{K\pi}
\right.\cr
&\left.
 - 4 \int_0^1 dx \Bigg[q^2 x (1 -x) \ell n \left({m_\pi^2 - q^2 x (1-x)\over
\mu^2}
\right)
- m_\pi^2 \ell n \left(1 - {q^2 x (1 - x)\over m_\pi^2} \right) \Bigg]
\right.\cr
& \left.
+ 2 q^2 \left( {m_K^2 - m_\pi^2  \over q^2 + 2 q \cdot (p_++p_-)} \right)
\left(  \phi_K(q^2) - \phi_\pi (q^2) + {1\over 6}\ell n\left({m_\pi^2\over
m_K^2}\right)
 \right) \ \ \  \right)
}\ \ \ \ .}
where
\eqn\anotherbigbas{\eqalign{
\phi_{K\eta} =   {2\over 9}  & (m_K^2   - m_\pi^2)^2  \left(
 \int_0^1 dy y \int_0^{1-y} dx \ {1\over \mu_1^2}
\right.\cr
&\left.
+ {1\over (q^2 + 2 q \cdot p_-)} \int_0^1 dx \ell n \left( 1 - {x (1 - x) (q^2
+ 2 q \cdot p_-)\over  m_K^2 (1-x) + m_\eta^2 x -m_\pi^2 x(1-x)  }\right)
\right)
\cr
+ {1\over 3} & (m_K^2  - m_\pi^2)  \left(
2 \int_0^1 dy \int_0^{1 - y} dx \ell n
\left(1 - {(q^2 x (1-x) + 2 q \cdot p_- x y)\over m_K^2 (1-y) + m_\eta^2 y
-m_\pi^2 y (1-y)}\right)
\right.\cr
&\left.
+ 3 \int_0^1 dy \int_0^{1-y} dx \ell n \left( 1 - {(q^2 x (1 -x) + 2q \cdot
p_ + x y)\over  m_K^2 (1-y) + m_\eta^2 y -m_\pi^2 y(1-y)}\right)
\right.\cr
&\left.
+  \int_0^1 dy y \int_0^{1-y} {1\over \mu^2_1} \Bigg[(q(1-x) + p_- y) \cdot
(4q+6 p_+ + 4p_-)
- 2 m_K^2 - 2 p_+ \cdot (q + p_-)\Bigg]
\right.\cr
&\left.
+  \int_0^1 dx \left(1 + x + {(x - 1) (3 m_K^2 - 2 m_\pi^2)\over q^2
+ 2 p_- \cdot q}\right)
\cdot \ell n \left(1 - {x (1 - x) (q^2 + 2q \cdot p_-)\over
m_K^2 (1-x) + m_\eta^2 x -m_\pi^2 x(1-x) }\right)
\right)
}\ \ \ }
with
\eqn\defmuone{
\mu_1^2 = m_K^2 (1 - y) + m_\eta^2 y - m_\pi^2 y (1 - y) - q^2 x (1 - x) - 2q
\cdot p_- x y\ \ \ \ .}
The other function is
\eqn\bigmoth{\eqalign{
\phi_{K\pi} = (m_K^2 - m_\pi^2) & \left(
 \int_0^1 dy \int_0^{1-y} dx \ell n \left(1 - {(q^2 x (1-x) + 2q \cdot p_- x y)
\over m_K^2 y + m_\pi^2 (1 - y)^2}
\right)
\right.\cr
&\left.
+ \int_0^1 dy \int_0^{1-y} dx \ell n \left( 1 - {q^2 x (1-x) + 2q \cdot p_+
x y  \over m_K^2 y + m_\pi^2 (1 - y)^2} \right)
\right.\cr
&\left.
+ 2 \int_0^1 dy y \int_0^{1-y} dx \  {(p_+ +p_-+q)\cdot (q(1-x) + y p_- - p_+)
\over \mu_2^2}
\right.\cr
&\left.
+ \int_0^1 dx \left( 1 + x + {m_K^2 (x - 1)\over q^2 + 2 q \cdot p_-} \right)
\ell n \left( 1 -  {x(1-x) (q^2 + 2 q \cdot p_-)\over m_\pi^2 (1 - x)^2 +
m_K^2 x} \right)
\right)
}\ \ \ \ }
with
\eqn\defmutwo{
\mu_2^2 = m_K^2 y + m_\pi^2 (1 - y)^2 - q^2 x (1 - x) - 2q \cdot p_- xy
 \ \ \ .}

The Gell-Mann--Okubo mass formula \gellmann\ has been used to simplify
some of the dependence on the pseudoscalar masses in \anotherbigbas\
and \bigmoth\ .
$F_+^{(2)}$ is obtained from \bigbas\  by taking $p_+ \rightarrow p_-$ and
$p_- \rightarrow p_+$.
Notice that the combination of coefficients $(a_1+2a_2)$ and $\lambda_{cr}$
that appear in the expression for
$F^{(2)}_\pm$ has a relative sign difference compared to the combination that
appears
in the expression for $f(s)$ given in eq.\kplusform .
The uncertainty in $ \lambda_{cr}(1 GeV)$ and $w_+$ gives rise to about
a $10\%$ uncertainty in the combination of counterterms that appears in \bigbas
{}.
The one photon part of the $K_L \rightarrow \pi^+ \pi^-
e^+ e^-$ decay amplitude is the largest and dominates the rate.  In the next
section we use the form factors calculated here to obtain
$d\Gamma( \K )/dq^2$.
One (scale independent) linear combination of counterterms
$w_L = a_3-a_4$ is not
determined by the  present experimental data and consequently
we cannot predict the rate for $\K$.
However, this is the only undetermined constant and the entire function
$d\Gamma( \K )/dq^2$ is experimentally accessible.

\newsec{ The Differential Decay Rate}

The $\K$ decay rate is obtained by squaring the invariant
matrix element (1.1), summing over the $e^+$ and $e^-$  spins, and
integrating over the phase space.  Since the $e^+$ and $e^-$ four
momenta
only occur in the lepton trace,
$Tr \left[ \k_- \gamma_\nu \k_+ \gamma_\mu \right]$, the
 phase space integrations over $k_-$ and $k_+$ produce a factor
\eqn\leptons{\eqalign{
\int{d^3 k_-\over (2\pi)^3 2k_-^0} \int{d^3 k_+\over (2\pi)^3 2k^0_+}  &
 (2\pi)^4 \delta^4 (q - k_- - k_+) Tr \left[ \k_- \gamma_\nu \k_+
\gamma_\mu\right]   \cr
& = {1\over 6\pi} (q_\mu q_\nu - q^2 q_{\mu \nu}) } \ \ \ \ .}
The remaining phase space integrations can be taken to be over $q^2$ and
the sum and difference of the pion energies in the $K_L$ rest frame,
$E_S = p_+^0 +
p_-^0, E_D = p_+^0 - p_-^0$.  The contribution of the form factors $F_\pm$
and $G$ to $d\Gamma/dq^2$ do not interfere.  Therefore, we can write
\eqn\diffrel{
{d\Gamma\over dq^2} (K_L \e) = {d\Gamma_G\over dq^2} \ \ + \ \
{d  \Gamma_F\over dq^2}\ \ \ ,}
where
\eqn\diffGF{\eqalign{
{d\Gamma_G\over dq^2} = {G_F^2 \alpha^2 s_1^2\over m_K f^2 2^6 (2\pi)^7 3q^2}
& \int
dE_S \int dE_D |G|^2 \cr
  \Bigg[ m_\pi^4 q^2 -  & m_\pi^2 (p_- \cdot q)^2 - m_\pi^2  (p_+\cdot  q)^2
+ 2 (p_+ \cdot p_-) (q \cdot p_+) (q \cdot p_-)-q^2(p_+\cdot  p_-)^2\Bigg]
\cr
{d\Gamma_F\over dq^2} = {G_F^2 \alpha^2 s_1^2\over m_K f^2 2^6 (2\pi)^7 3q^4}
&
\int\ dE_S \int dE_D \Bigg[
| F_+ \ q \cdot p_+ + F_- \  q \cdot p_-|^2
\cr
&
 - q^2 \left( |F_+|^2 m_\pi^2 + |F_-|^2 m_\pi^2 + 2 Re (F_+ F_-^*) p_+\cdot p_-
\right) \Bigg]
}\ \ \ .}
In eq. \diffrel\  and eq. \diffGF\  the difference of pion energies is
integrated over
the region \ \ \ $-~E_D^{(max)}~<~E_D~<~E_D^{(max)}$ where
\eqn\pionenediff{
E_D^{(max)} = \sqrt{  2 m_K E_S + q^2 - m_K^2 - 4m_\pi^2 \over
2 m_K E_S + q^2 - m_K^2}
\sqrt{(m_K - E_S)^2 - q^2}
\ \ \ ,}
and the sum of pion energies is integrated over the region
$E_S^{(min)} < E_S < E_S^{(max)}$ where the boundaries are
\eqn\pionenesum{\eqalign{
E_S^{(max)} & = m_K - \sqrt{q^2}  \cr
E_S^{(min)} & = {m_K^2 - q^2 + 4 m_\pi^2\over 2m_K}
}\ \ \ .}
The scalar  products appearing in the expression for the rates are easily
expressed in
terms of $E_S, E_D$ and $q^2$:
\eqn\ratedotprods{\eqalign{
p_+ \cdot p_-  & = {1\over 2} (q^2 - m_K^2 - 2 m_\pi^2 + 2 m_K E_S) \cr
q \cdot p_+  & = {1\over 2}(- m_K E_S + m_K E_D - q^2 + m_K^2) \cr
q \cdot p_- & = {1\over 2}(- m_K E_S - m_K E_D - q^2 + m_K^2)
}\ \ \ \ .}
The form factors $F_\pm^{(1)}$ and $F_\pm^{(2)}$
have the opposite property under interchange of pion momenta and
consequently
they do not interfere in $d\Gamma/dq^2$.
Neglecting terms in chiral expansion of ${\cal O}(p^6)$ and higher the
differential decay rate given in \diffrel\ becomes
\eqn\diffFFG{
{d\Gamma \over dq^2} (K_L \e) =
{d\Gamma_{G^{(2)}}\over dq^2} + {d \Gamma_{F^{(1)}}\over dq^2} +
{d\Gamma_{F^{(2)}}\over dq^2}\ \ \ .}

\noindent In
\fig\rateplot{The differential decay spectrum as a function of $y$
the invariant mass of the lepton pair normalized to $m_K-2 m_\pi$.
The dot-dashed curve is the contribution from $F_\pm^{(1)}$,
the dotted curve is the contribution from $F_\pm^{(2)}$ with $w_L=0$
and the dashed curve is the contribution from $G^{(2)}$.
The total differential decay rate for $w_L=0$ is given by the solid curve.}
we have graphed for each of the three terms on the
right hand side of \diffFFG\

\eqn\vardiff{
{1\over\Gamma_{K_L}}  {d\Gamma\over dy} = 2y~(m_K-2m_{\pi})^2
{1\over\Gamma_{K_L}}  {d\Gamma \over dq^2}\ \ \ ,}
 where $y=\sqrt{q^2}/(m_K-2m_{\pi})$, $\Gamma_{K_L}$ is the total width of the
$K_L$ and we have set $w_L=0$.

Integrating the three terms on the rhs of \diffFFG\ over  the invariant mass
interval $q^2 > (30MeV)^2$ (corresponding to $y > 0.14$) we find that for
$w_L=0$
\eqn\minthirty{
10^8 \cdot Br (\K ; q^2 > (30 MeV)^2  ) \ =\  3.8 \ +\  0.78 \ + \  3.4 \ =\
8.0 \ \ \
\ . }
The branching fraction over this range of $e^+e^-$ invariant mass is dominated
by the region of  low $q^2$ and for typical values of $w_L$ it
receives comparable contributions from the form factors  $G$ and $F^{(2)}$.
However, in the region of high $q^2$ the branching fraction is likely to be
dominated  by the $F_\pm^{(2)}$ form factor.
For $q^2 > (80MeV)^2$ (corresponding to $y > 0.37$) and $w_L=0$
the three terms on the rhs of \diffFFG\  contribute
\eqn\mineighty{
10^8 \cdot Br (K_L \e; q^2 > (80MeV)^2) \  =\  0.61 \ +\  0.07 \ + \  1.9 \ =\
2.6\ \ \ \
.}

A summary of our results for the rate can be found in Table 1.
We have displayed the contribution to the branching ratio (in units of
$10^{-8}$)
from  the three form factors
$G$, $F^{(1)}$ and $F^{(2)}$ for different values of the minimum lepton pair
invariant mass $q^2_{\rm min}$.
Since the  loop contribution to the form factor $F_\pm^{(2)}$ is small,
it will be difficult to extract a unique value for $w_L$ from $d\Gamma/dq^2$
data alone; a two-fold ambiguity in the value of $w_L$ will persist.

\bigskip\bigskip
\begintable
Lower cut $q^2_{\rm min}$|~~ $Br(10^{-8})_G$~~ |~~ $Br(10^{-8})_{F^{(1)}}$~~|~~
$Br(10^{-8})_{F^{(2)}}$  \cr
$ ( 10 {\rm MeV} ) ^2 $| $  8.8  $ | $3.3$ | $3.6 - 3.4 w_L + 0.8w_L^2$
\crnorule
$ ( 20 {\rm MeV} ) ^2 $| $  5.6  $ | $1.5$ | $3.5 - 3.3w_L + 0.8w_L^2$
\crnorule
$ ( 30 {\rm MeV} ) ^2 $| $  3.8  $ | $0.8$ | $3.4 - 3.2w_L + 0.8w_L^2$
\crnorule
$ ( 40 {\rm MeV} ) ^2 $| $  2.7  $ | $0.5$ | $3.1- 3.0w_L + 0.7w_L^2$ \crnorule
$ ( 60 {\rm MeV} ) ^2 $| $  1.3  $ | $0.2$ | $2.6 - 2.4w_L + 0.6w_L^2$
\crnorule
$ ( 80 {\rm MeV} ) ^2 $| $  0.6  $ | $0.07$ | $1.9-1.8w_L + 0.4w_L^2$ \crnorule
$ ( 100 {\rm MeV} ) ^2 $| $  0.3  $ | $0.03$ | $1.3-1.2w_L + 0.3w_L^2$
\crnorule
$ ( 120 {\rm MeV} ) ^2 $| $  0.1  $ | $0.01$ | $0.74-0.68w_L+0.16w_L^2$
\crnorule
$ ( 180 {\rm MeV} ) ^2 $| $  0.00072  $ | $0.0001$ | $0.027 - 0.025w_L +
0.006w_L^2$
\endtable
\bigskip
\centerline{{\bf Table 1: Contributions to the Branching Ratio $(10^{-8})$
for a range of $q^2_{\rm min}$ }}
\bigskip

\newsec{ The $Z$-penguin and $W$-box Amplitude}

The short distance $W$-box and $Z$-penguin diagrams give the effective
Lagrange density
\eqn\shortlag{
\L_{SD} = \xi\ {s_1 G_F \alpha\over\sqrt{2}} \
\bar s \gamma_\mu (1 - \gamma_5)
d\   \bar e \gamma^\mu \gamma_5 e \ \ +\ \ h.c.\ \ \ .}
Here we only keep the part that contains the lepton axial current (the
vector current is neglected).  It is only the axial current that gives
rise to observables that are antisymmetric  under interchange of $e^+$ and
$e^-$
 four momenta, $k_+\leftrightarrow k_-$.

In \shortlag\ the quantity $\xi$ receives significant contributions from both
the
top quark
and charm quark loops and is given by
\eqn\xiexpress{
\xi = - \tilde{\xi}_c + \left( {V_{ts}^* V_{td}\over V_{us}^* V_{ud}} \right)
\tilde{\xi}_t \ \ \ \ ,}
where
\eqn\xicomps{
 \tilde{\xi}_q = \tilde{\xi}_q^{(Z)}\  +\  \tilde{\xi}_q^{(W)} \ \ \ ,}
is the sum of the contributions of the $Z$-penguin and $W$-box diagrams.  It is
convenient to express the combination of elements of the
Cabibbo--Kobayashi--Maskawa matrix that enters in $\xi$ in terms of $|V_{cb}|$
and the standard coordinates $\rho +i\eta$ of the unitarity triangle
\eqn\vbcrel{
V_{ts}^* V_{td} / V_{us}^* V_{ud} = (\rho - 1 + i\eta)|V_{cb}|^2\ \ \ .}
A  value of $|V_{cb}| \simeq 0.04$ is obtained from inclusive $B \rightarrow
X_c
e\bar\nu_e$ decay and from exclusive $B \rightarrow D^* e \bar\nu_e$ decay.
Although the values of $\rho$ and $\eta$ are not determined by present data,
they are expected to be of order unity.

The quantities $\tilde \xi_c$ and $\tilde \xi_t$ have been calculated
including perturbative QCD corrections at the next to leading logarithmic
level
\ref\bubua{G. Buchalla and A.J. Buras, \np{398}{1993}{285}.}
\ref\bubub{G. Buchalla and A.J. Buras, \np{400}{1993}{225}.}.
There is some sensitivity to the values of $\Lambda_{QCD}$~,~$m_c$ and
$m_t$ but $\tilde \xi_c$ is of order $10^{-4}$ and $\tilde \xi_t$ is of order
unity.

The quark-level Lagrange density in eq. (4.3) can be converted into a Lagrange
density involving the $\pi, K$ and $\eta$ hadrons using the Noether procedure.
Equating the QCD chiral currents with those obtained from chiral
variations of the effective lagrangian in eq. \lagpi\  leads to
\eqn\currentshort{
\L_{SD} = - \xi \ {iG_F \alpha s_1\over 2\sqrt{2}} \  f^2  Tr (\partial^\mu
\Sigma \Sigma^\dagger T) \ \  \bar e \gamma_\mu \gamma_5 e \ +\  h.c. \ \ \ \
.}
Expanding out $\Sigma$ in terms of  the meson fields $M$ we find that the
Lagrange density
\currentshort\
implies that the short  distance contribution to the $K_L \e$ decay  amplitude
from the $W$-box and $Z$-penguin diagrams is
\eqn\SDthreepi{
M^{(SD)} =  { G_F s_1 \alpha\over f} \left(  \xi\  p_-^\mu + \xi^*\  p_+^\mu
\right)
\ \ \bar u (k_-) \gamma_\mu \gamma_5 v(k_+) \ \ \ .}

\newsec{The Asymmetry $A_{CP}$}

It is the interference of $M^{(SD)}$ in \SDthreepi\  with $M^{(1\gamma)}$ in
\onegam\
that produces the asymmetry $A_{CP}$ defined in \cpvar\ .
For   calculation of $A_{CP}$ it is convenient to use the phase space variables
used by Pais and Treiman
\ref\paistr{A. Pais and S. Trieman, \physrev{168}{1968}{1858}.}
for $K_{\ell 4}$ decay (rather than those used
for the total rate in Section 3).
They are: $q^2 = (k_+ + k_-)^2;~ s = (p_+ + p_-)^2;~ \theta_\pi$
the angle formed by the $\pi^+$ three momentum and the
$K_L$ three-momentum in the $\pi^+ \pi^-$ rest frame; $\theta_\ell$, the angle
between the $e^-$ three momentum and the $K_L$ three momentum in the
$e^+ e^-$
rest frame; $\phi$, the angle between the normals to the planes defined in the
$K_L$ rest frame by the $\pi^+ \pi^-$ pair and the $e^+ e^-$ pair.
In terms of these variables
\eqn\asymmvar{
{ \PP \over |\PP|} =  {\rm sign}  (\sin \phi)  \ \ \ ,}
and the asymmetry is
\eqn\asymmexp{
A_{CP} = {1\over 2^7 (2\pi)^6 m_K^3 \Gamma_{K_L}}  \left(\int_0^{2\pi} d\phi\
{\rm sign}(\sin\phi)\right)
 \int dc_\pi\  dc_e\  ds\  dq^2
\beta \ X \ Re \left( M^{(SD)^{*}} M^{(1\gamma))}\right)  \ \ \ ,}
where $c_\pi = \cos\theta_\pi$, $c_e = \cos\theta_e$.
The other kinematic functions appearing in this expression are
\eqn\kinasym{\eqalign{
\beta & = [1 - 4m_\pi^2/s]^{1/2}\cr
X & = \left[\left({m_K^2 - s - q^2\over 2} \right)^2 - sq^2\right]^{1/2} }\ \ \
.}
In order to evaluate the contributing form factors the following scalar
products of
four vectors are required:
\eqn\dotprods{\eqalign{
q\cdot p_+ & = {1\over 4} (m_K^2 - s - q^2) - {1\over 2}\beta X\cos\theta_\pi
\cr
q\cdot p_- & = {1\over 4} (m_K^2 - s - q^2) + {1\over 2}\beta X\cos\theta_\pi
\cr
p_+\cdot p_- & = {1\over 2} ( s-2 m_\pi^2)\cr
\varepsilon_{\alpha\beta\sigma\rho} p_+^\alpha p_-^\beta k_+^\sigma k_-^\rho
& = -{1\over 4}\beta X \sqrt{s q^2} \sin\theta_e \sin\theta_\pi \sin\phi
}\ \ \ .}
If the variables $s$ and $q^2$ are not
integrated over the complete phase space then it is understood that the same
is to be done for the $K_L$ width $\Gamma_{K_L}$ in the denominator of
eq.~\asymmexp .

The form factor $G$ does not enter into
$Re\left( M^{(SD)^{*}} M^{(1\gamma)}\right) $
(a sum over $e^+$ and $e^-$  spins is understood).  Integrating our $\cos
\theta_e$ and $\phi$ we find that
\eqn\asymmint{\eqalign{
A_{CP} = {G_F^2 s_1^2 \alpha^2\over 2^8 (2\pi)^6 f^2 m_K^3 \Gamma_{K_L}} &
\int d c_\pi\  ds\  dq^2\  \sin\theta_\pi \ \cr
\beta^2\  X^2 \  &  \sqrt{s \over q^2}
\Bigg[Im(\xi)  \left(  Re(F_+) + Re(F_-) \right)
+ Re(\xi) \left(Im( F_+) - Im( F_-)  \right)\bigg]  }\ \ \ .}
The integration over $\cos\theta_\pi$ implies that at leading non-trivial order
of
chiral perturbation theory $Im( F_+)  - Im( F_-)
\rightarrow Im( F_+^{(1)} ) - Im( F_-^{(1)}) $ reflecting indirect CP violation
from $\epsilon$
and $Re(F_+) + Re(F_-) \rightarrow Re( F_+^{(2)}) + Re(F_-^{(2)})$ in
eq.\asymmint\ .

Using \xiexpress\  and \vbcrel\  we can write the CP violating asymmetry in
terms
of the  real and imaginary parts of the CKM elements
\eqn\cpckm{
A_{CP} = A_1 \left(  (\rho - 1)|V_{cb}|^2 \tilde{\xi}_t - \tilde{\xi_c}\right)
- A_2 \eta |V_{cb}|^2 \tilde{\xi_t}\ \ \ ,}
where $A_1$ arises from indirect CP nonconservation (i.e. $\overline{K}^0-K^0$
mixing )
and $A_2$ arises from direct CP nonconservation.
We are only able to predict $|A_{CP}|$ since the sign of
$g_8$ is not known.
Our expressions for $F_\pm^{(1)}$ and $F_\pm^{(2)}$ with $w_L=0$ give
(up to an overall sign)
\eqn\ainumsa{
A_1 = 2.7\times 10^{-2} \qquad \qquad, \qquad A_2 = 3.9\times 10^{-2}\ \ \ ,}
for $ q^2 \geq (30MeV)^2 $ and
\eqn\ainumsb{
A_1  = 2.4\times 10^{-2}\qquad \qquad, \qquad A_2 = 8.4\times 10^{-2} \ \ \ ,}
for $ q^2 \geq (80 MeV)^2$.
In Table 2 we present  $A_1$ and $A_2$ for a range of values of the
minimum lepton pair invariant mass, $q^2_{\rm min}$, normalized to the
branching
ratios given in Table 1 assuming $w_L=0$.

\bigskip\bigskip
\begintable
Lower cut $q^2_{\rm min}$|~~ $A_1$~~ |~~ $A_2$~~   \cr
$ ( 10 {\rm MeV} ) ^2 $ | $2.0 \times 10^{-2}$  | $  2.0\times 10^{-2}
$\crnorule
$ ( 20 {\rm MeV} ) ^2 $ | $2.5 \times 10^{-2}$  | $  3.0\times 10^{-2}
$\crnorule
$ ( 30 {\rm MeV} ) ^2 $ | $2.7 \times 10^{-2}$ | $  3.9\times 10^{-2}
$\crnorule
$ ( 40 {\rm MeV} ) ^2 $ | $2.8 \times 10^{-2}$  | $  4.8\times 10^{-2}
$\crnorule
$ ( 60 {\rm MeV} ) ^2 $| $2.7 \times 10^{-2}$  | $  6.8\times 10^{-2}  $
\crnorule
$ ( 80 {\rm MeV} ) ^2 $| $2.4 \times 10^{-2}$  | $  8.4\times 10^{-2}  $
\crnorule
$ ( 100 {\rm MeV} ) ^2 $ | $2.1 \times 10^{-2}$  | $  9.8\times 10^{-2}
$\crnorule
$ ( 120 {\rm MeV} ) ^2 $| $1.8 \times 10^{-2}$  | $  0.11  $ \crnorule
$ ( 180 {\rm MeV} ) ^2 $| $1.3 \times 10^{-2}$  | $  0.13  $
\endtable
\bigskip
\centerline{{\bf Table 2: The CP violating quantities $A_1$,  $A_2$  with
$w_L=0$
for different  values of $q^2_{\rm min}$ }}
\bigskip

\noindent We find that direct and indirect sources of CP nonconservation give
comparable
contributions to $A_{CP}$.
In our computation we have neglected final state $ \pi \pi $ interactions which
are
formally higher order in chiral perturbation theory.
With the values of $A_1$ and  $A_2$ given in  Table 2, $|A_{CP}|$ is
only of order $10^{-4}$ and further refinements of our calculation do not seem
warranted.

\newsec{The Asymmetry $B_{CP}$}

Using the kinematic variables introduced in the previous
section the CP violating observable $B_{CP}$ is defined as
\eqn\bcp{
B_{CP} = < {\rm sign} (\sin\phi\cos\phi) > \ \ \ .}
At leading order in chiral perturbation theory it arises
from the interference of $F^{(1)}_{\pm}$ with $G^{(2)}$.
The CP violating form factors $F^{(1)}_{\pm}$
are not small because they occur at a lower order in chiral perturbation theory
than the other form factors, $F_\pm^{(2)}$ and $G^{(2)}$.
Consequently, as was noted in  refs. \sehwan\  and \heilseh ,
$B_{CP}$ is quite large.
Neglecting $M^{(SD)}$ we find
after integrating over $\phi$ and $cos\theta_e$ that
\eqn\bcpform{
B_{CP} = {G_F^2 s_1^2 \alpha^2\over 3\  2^7 (2\pi)^8 f^2 m_K^3 \Gamma_{K_L}}
\int d c_\pi\  ds\  dq^2\  \sin^2\theta_\pi  \
\beta^3\  X^2 \ {s \over q^2}  Im \left[G \ (  F_+^* - F_-^* )\right]  \ \ \ .}
If the variables $s$ and $q^2$ are not integrated over the entire phase space
then it is understood that the same is to be done to the $K_L$ width
$\Gamma_{K_L}$ in the denominator of \bcpform .
The form factor $G$ is real at leading order in chiral perturbation theory and
the imaginary part
arises from the phase in $F_+-F_-$ induced by $K^0-\overline{K}^0$ mixing.
The integration over $\cos\theta_\pi$ implies that
$F_+-F_-\rightarrow F_+^{(1)}-F_-^{(1)}$ in eq.\bcpform\ .
Using our expressions for  $F^{(1)}_{\pm}$ and the value of $|G^{(2)}|$ we
find that with $w_L=0$ $|B_{CP}|\simeq 6.3\%$ for  $q^2\ > \ (30\ MeV)^2$ and
$|B_{CP}|\simeq 2.4\%$ for $q^2\ >\ (80\ MeV)^2$.
The asymmetry for a range of  values of $q^2_{\rm min}$ are shown in Table 3.
\bigskip\bigskip
\begintable
Lower cut $q^2_{\rm min}$ |~~ $  \vert B_{CP}\cdot Br(10^{-8}) \vert   $ (\%)
{}~~   \cr
$ ( 10 {\rm MeV} ) ^2 $| $ 134  $  \crnorule
$ ( 20 {\rm MeV} ) ^2 $| $ 78  $  \crnorule
$ ( 30 {\rm MeV} ) ^2 $| $ 50  $  \crnorule
$ ( 40 {\rm MeV} ) ^2 $| $ 33  $  \crnorule
$ ( 60 {\rm MeV} ) ^2 $| $ 14  $  \crnorule
$ ( 80 {\rm MeV} ) ^2 $| $ 6.3  $  \crnorule
$ ( 100 {\rm MeV} ) ^2 $| $ 2.5  $  \crnorule
$ ( 120 {\rm MeV} ) ^2 $| $ 0.92  $  \crnorule
$ ( 180 {\rm MeV} ) ^2 $| $0.0086  $
\endtable
\bigskip
\centerline{{\bf Table 3: The CP violating observable $|B_{CP}\cdot Br(10^{-8})
 |$
for a range of values of $q^2_{\rm min}$ }}
\bigskip
\noindent  Note that in Table 3 $ Br(10^{-8}) $ denotes the $\K$ branching
ratio
in units of $10^{-8}$ with the same  cut on $q^2$ imposed.
We have neglected final state $\pi \pi $ interactions because they arise at
higher
order in chiral perturbation theory.
Our prediction for $|B_{CP}|$ has considerable uncertainty because of the
neglect of final state $\pi \pi $ interactions and because neglected
${\cal O}(p^6)$  contributions to $G$ seem to be important.

\newsec{ Conclusions}

In this paper we have calculated the one-photon contribution to the $\K$
decay rate.  We used chiral perturbation theory to determine the form factors
and
for $e^+e^-$ pairs with high invariant mass
( $q^2 >> 4 m_e^2$)  found important new contributions that were not included
in
previous work \sehwan\heilseh\ .
The amplitude for $\K$ depends on the undetermined (renormalization scale
independent)
combination of counterterms $w_L$.
We found that for $q^2 = (k_+ + k_-)^2 > ( 30\  MeV)^2$ the branching ratio for
$\K$
is approximately
$ ( 8.0 - 3.2 w_L + 0.8w_L^2)  \times 10^{-8}$
and for  $q^2  > ( 80\  MeV)^2$ the branching ratio is
approximately
$ (2.6 - 1.8 w_L + 0.4 w_L^2) \times 10^{-8}$.

One interesting aspect of this decay mode is that the CP even
component of the $K_L$ state  contributes at a lower order in
chiral perturbation theory than the CP odd component.
This enhances CP violating effects in $\K$ decay.
For example, the CP violating observable \sehwan\ \heilseh\
$B_{CP} = <{\rm sign} (\sin\phi \cos\phi)>$
, where $\phi$ is the angle between the normals to the $\pi^+\pi^-$ and
$e^+e^-$ planes,  is about $6\%$ for $q^2 > (30 MeV)^2$ if $w_L=0$.
The CP violating observable $A_{CP}= <{\rm sign} (\sin\phi )>$
arises from the interference of
W-box and Z-penguin amplitudes with the one-photon part of the decay amplitude.
Unfortunately,  we find that $A_{CP}$ is of order $10^{-4}$ and hence most
likely
unmeasurable.

Chiral Perturbation theory has been extensively applied to nonleptonic,
semileptonic and
radiative kaon decays.
The study of $\K$ offers an opportunity to determine the linear combination
of coefficients in the ${\cal O}(p^4)$ chiral lagrangian that we call
$w_L$ and to test the applicability of ${\cal O}(p^4)$ chiral perturbation
theory
for kaon decay.

Some improvements in our calculations are possible.   While a full computation
of the ${\cal O}(p^6)$ contribution to $F_\pm$ and $G$ arising from two-loop
diagrams and new local operators does not seem feasible it should be
possible to calculate the leading contribution to the absorptive parts of
$G$ and  $F_+-F_-$.
Note that the absorptive parts come from both $\pi\pi\rightarrow\pi\pi$
rescattering
and because of CP nonconservation from $\pi\pi\rightarrow\pi\pi\gamma$.
We hope to present results for this in a future publication.

\bigskip\bigskip\bigskip
\centerline{{\bf Acknowledgements}}
\bigskip
MJS would like to thank the High Energy Physics group at Caltech
for kind hospitality during part of this work.

\listrefs
\listfigs
\vfill\eject

\insertfigmed{Figure 1}{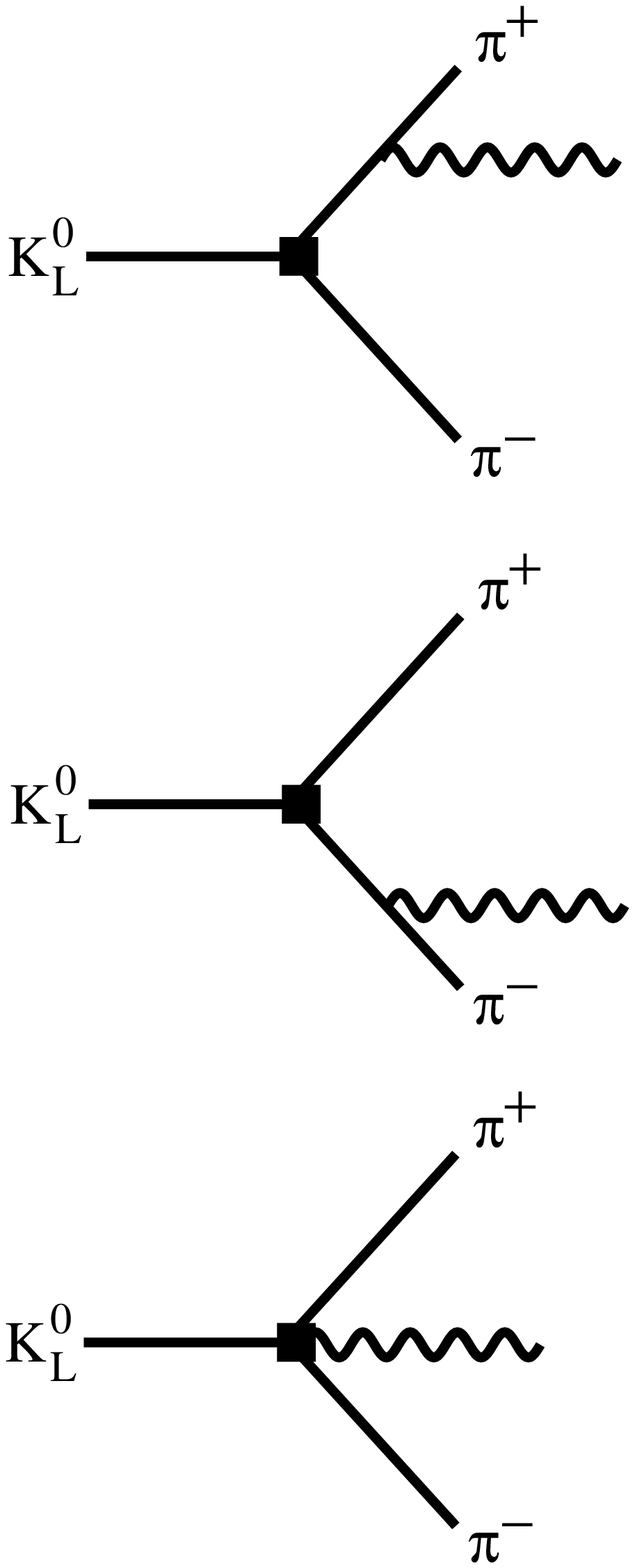}
\insertfig{Figure 2}{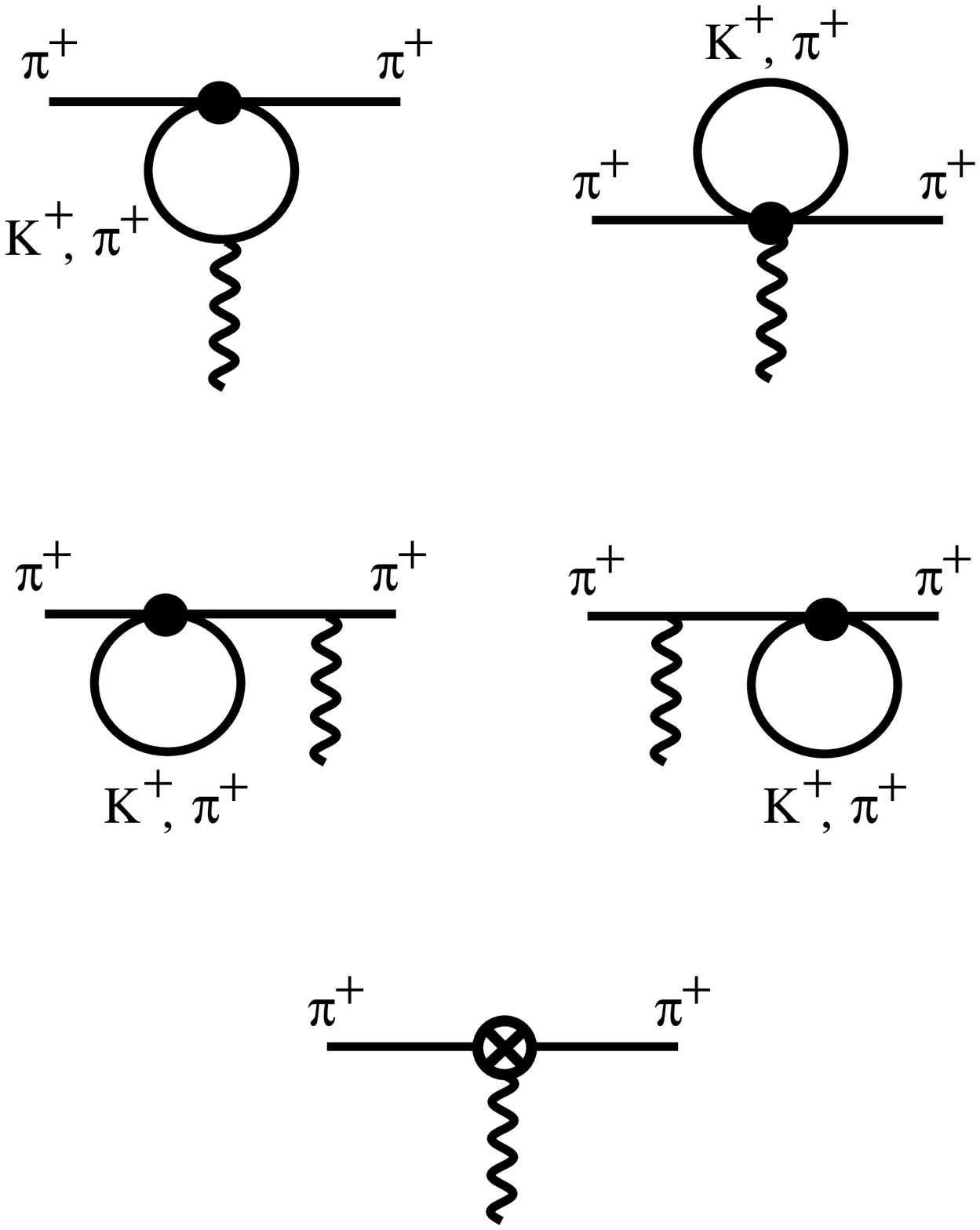}
\insertfig{Figure 3(a)}{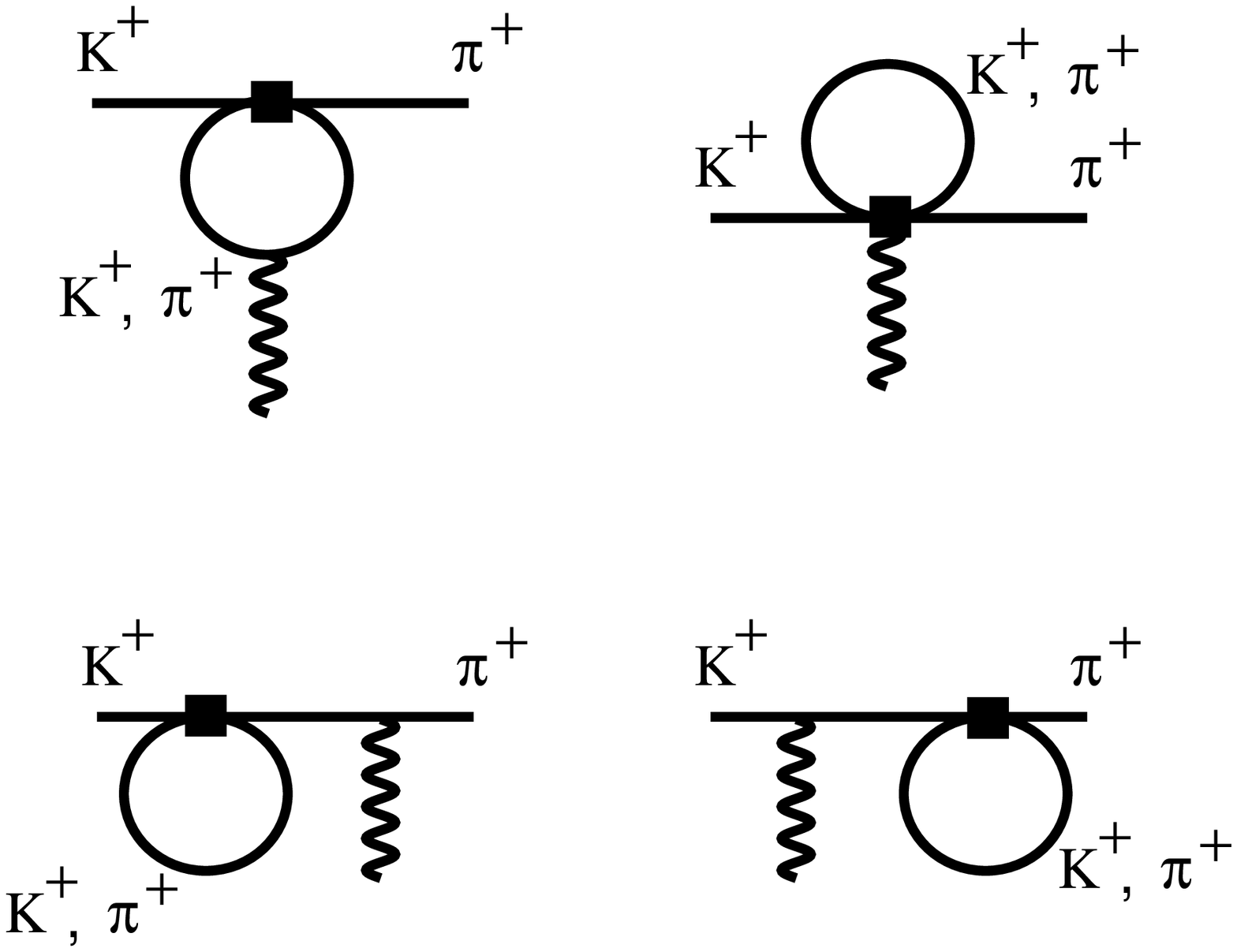}
\insertfig{Figure 3(b)}{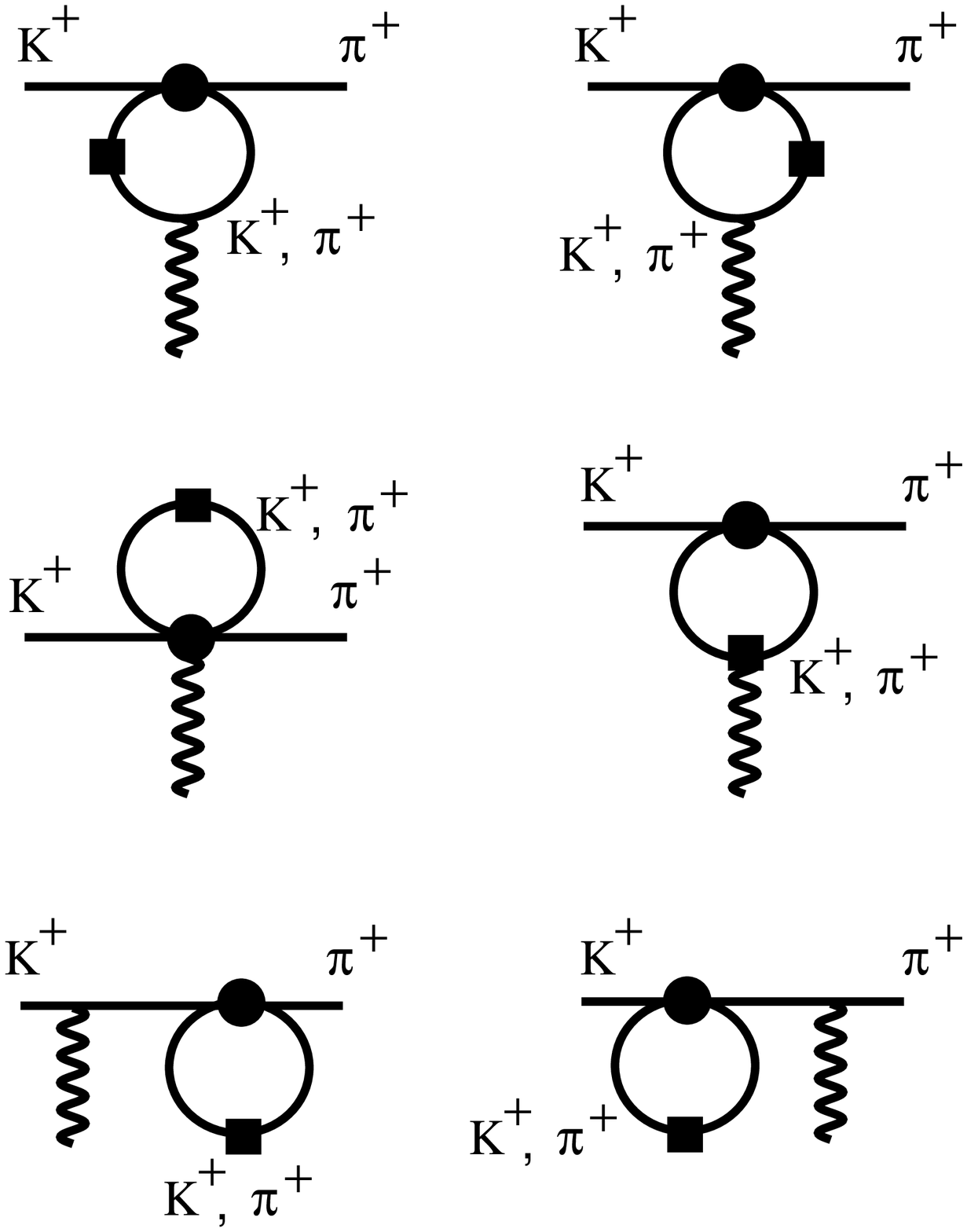}
\insertfig{Figure 3(c)}{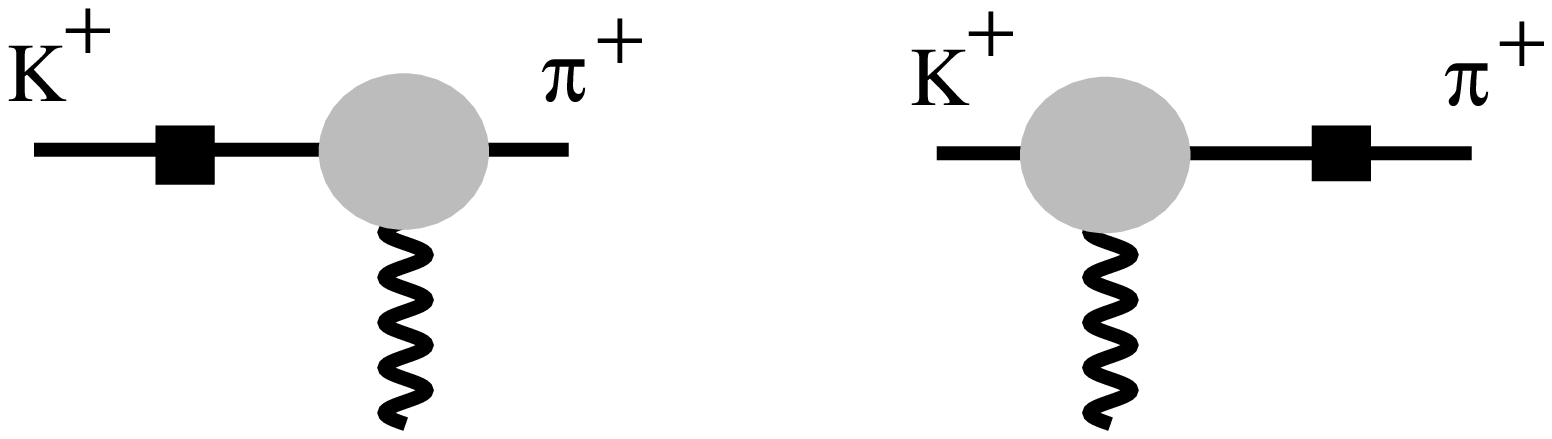}
\insertfigmed{Figure 3(d)}{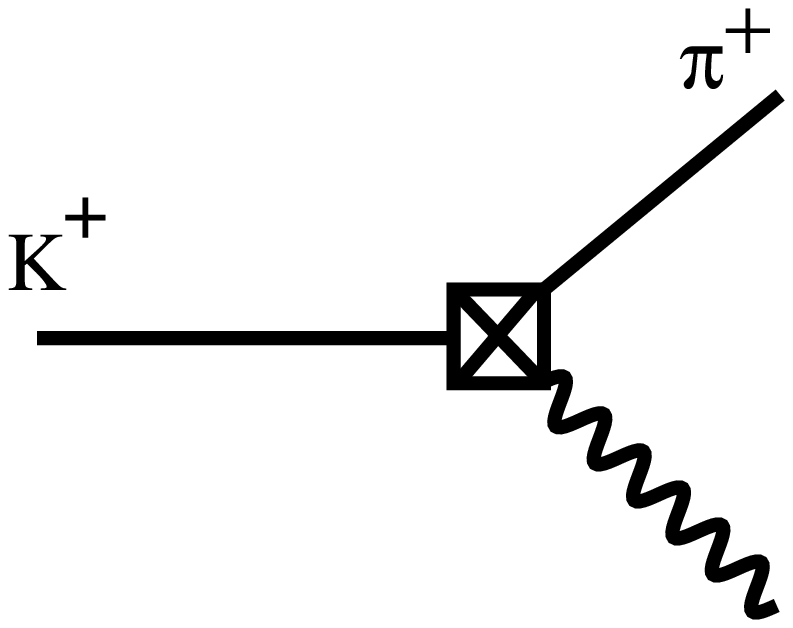}
\insertfig{Figure 4(a)}{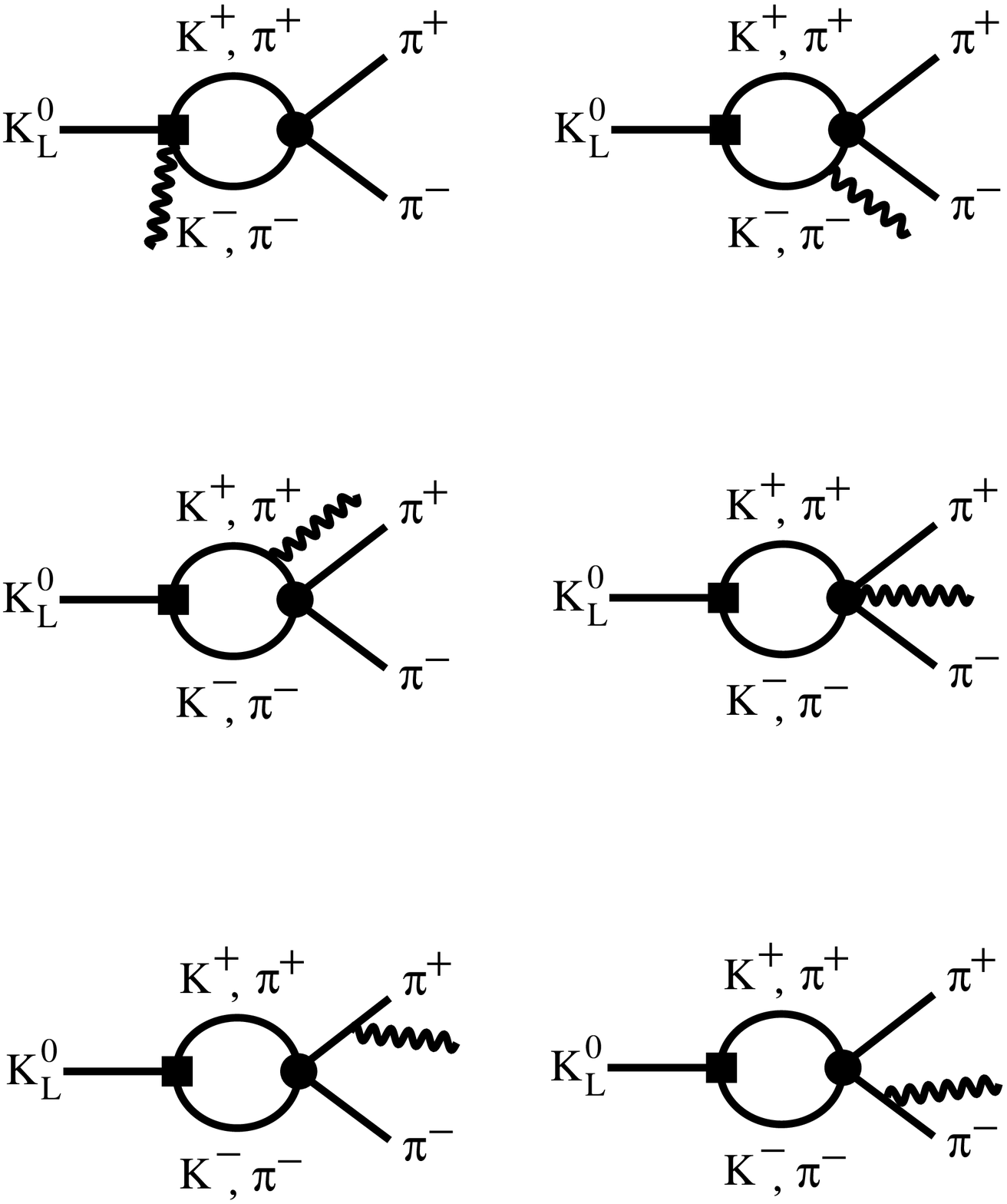}
\insertfigmedbig{Figure 4(b)}{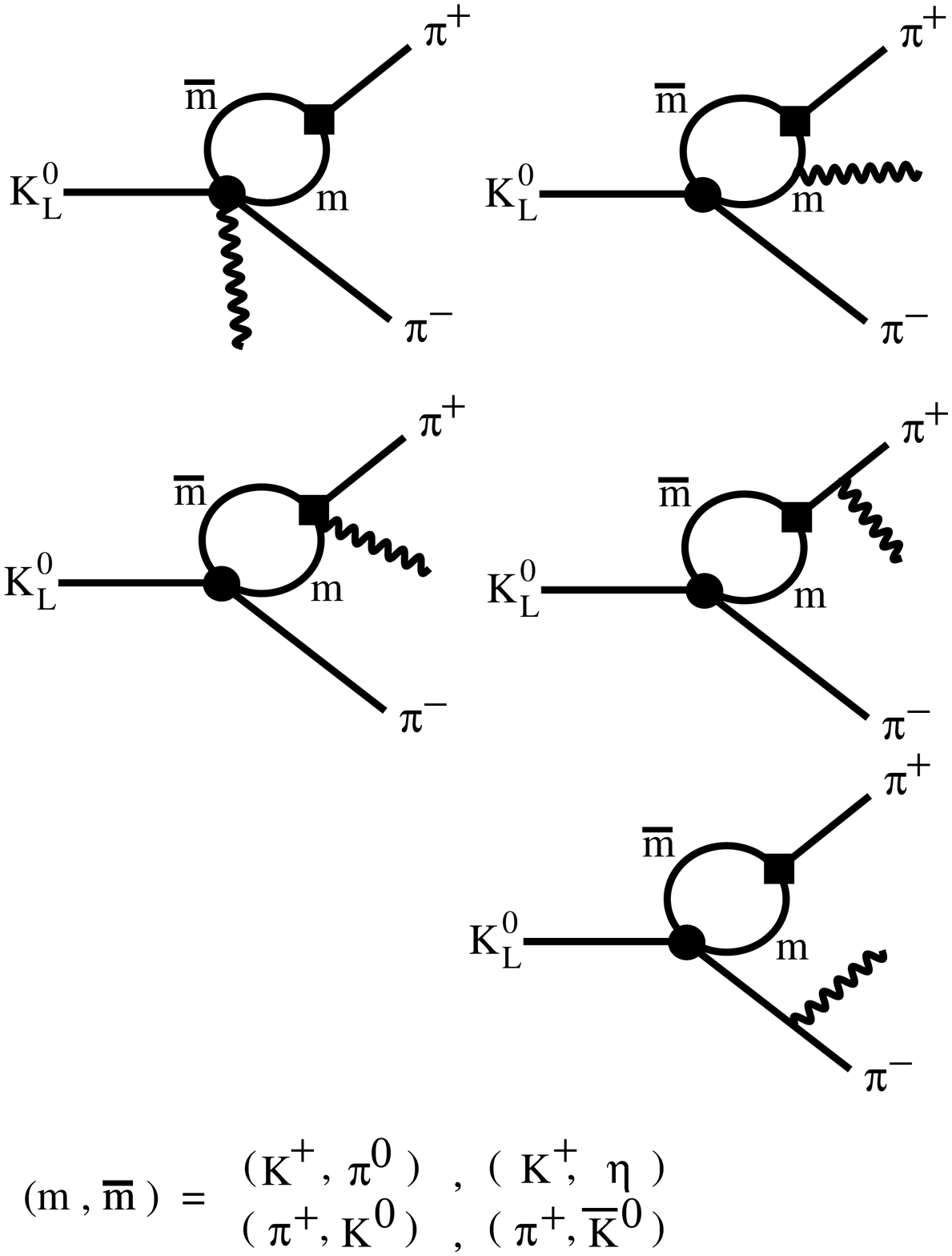}
\insertfigmedbig{Figure 4(c)}{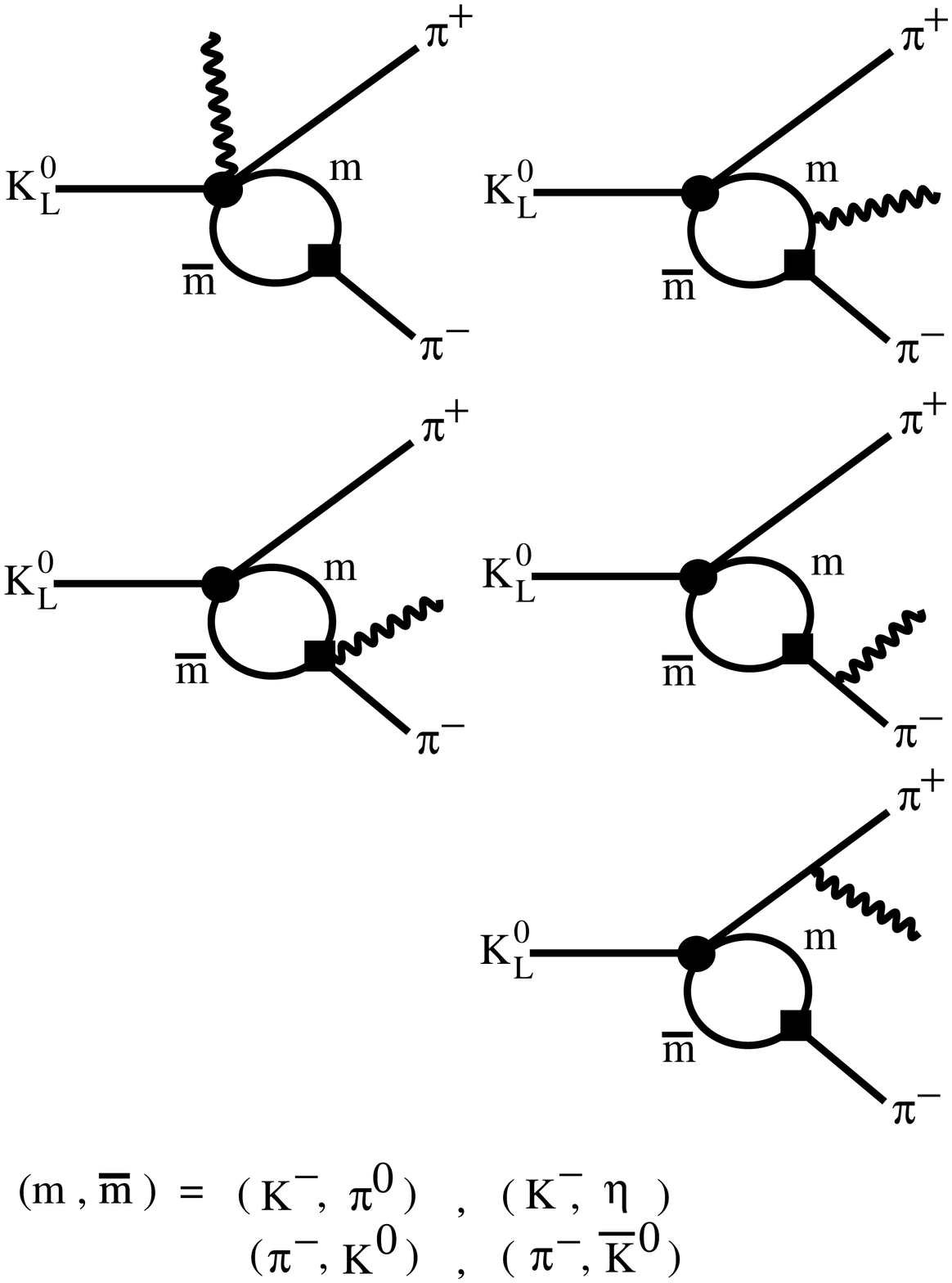}
\insertfig{Figure 4(d)}{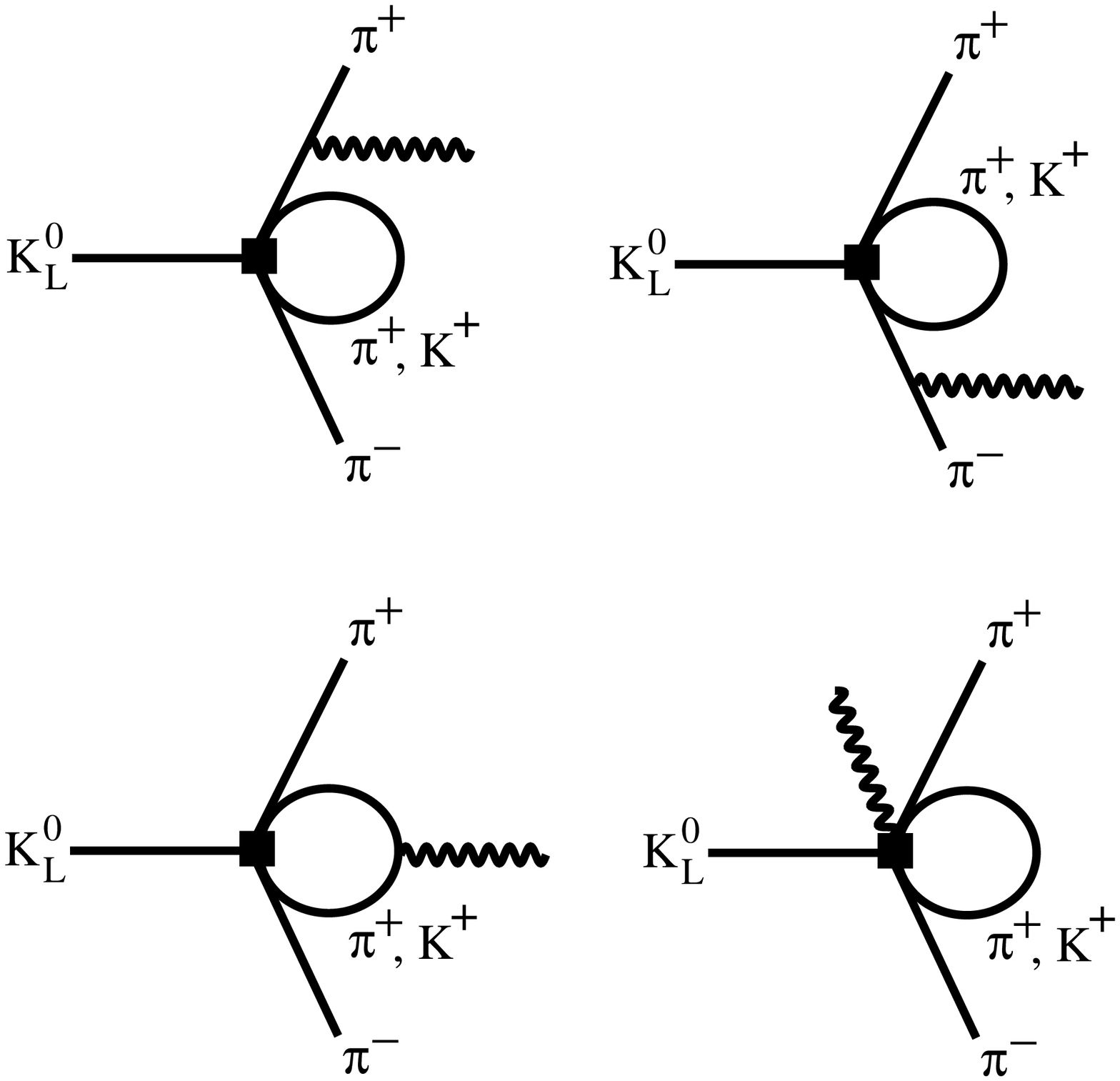}
\insertfig{Figure 4(e)}{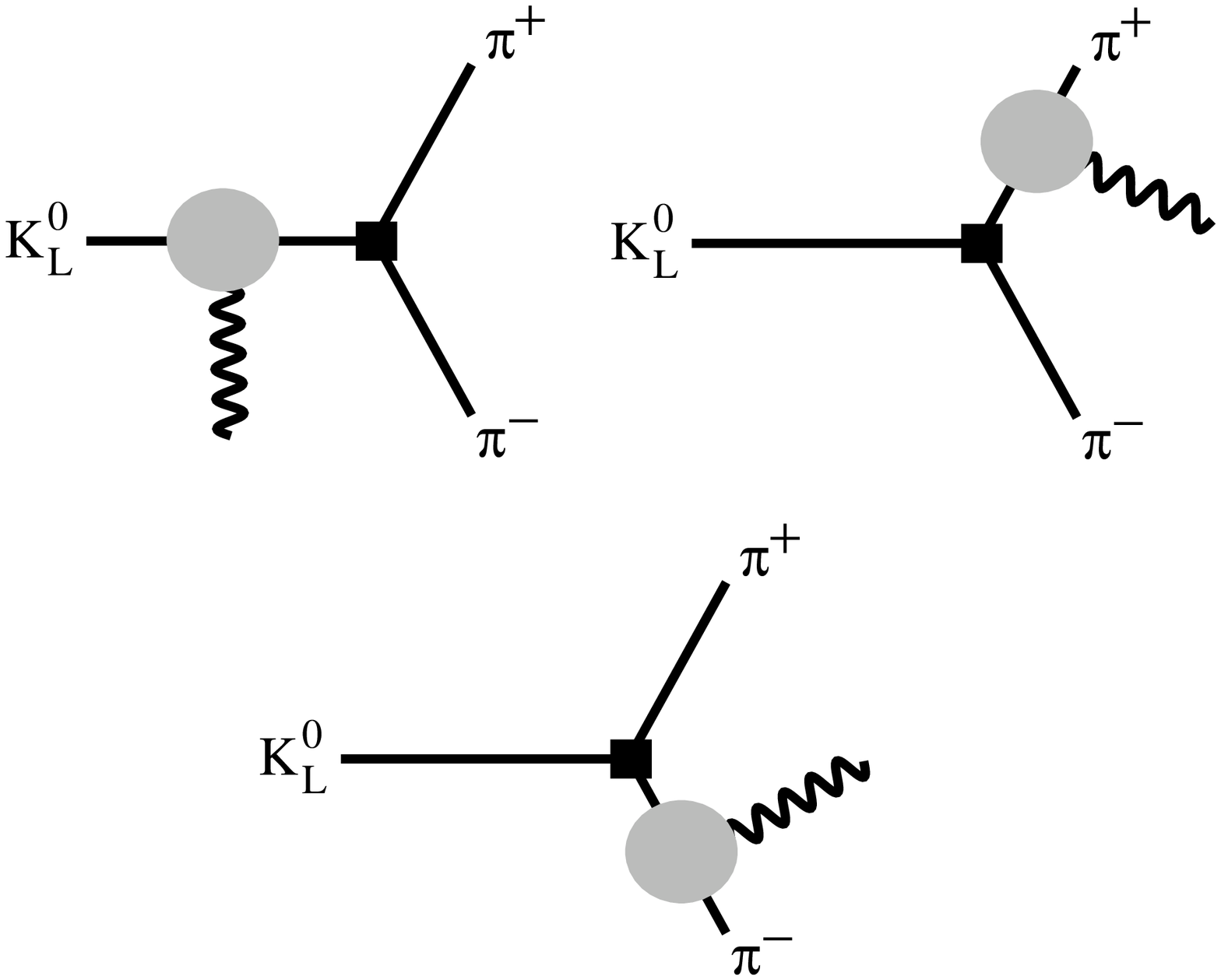}
\insertfigmed{Figure 4(f)}{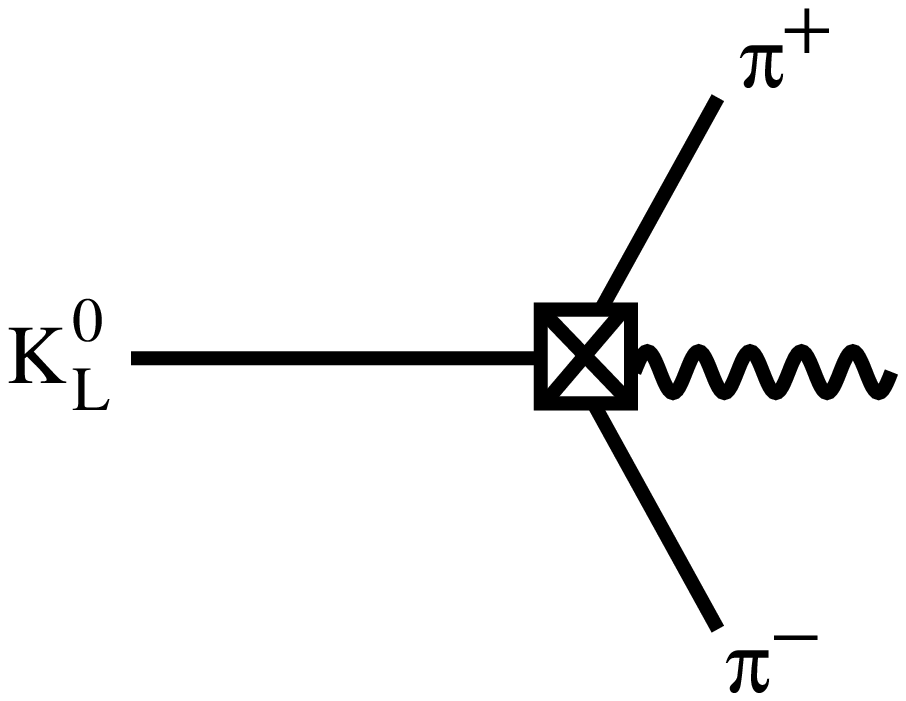}
\insertfig{Figure 5}{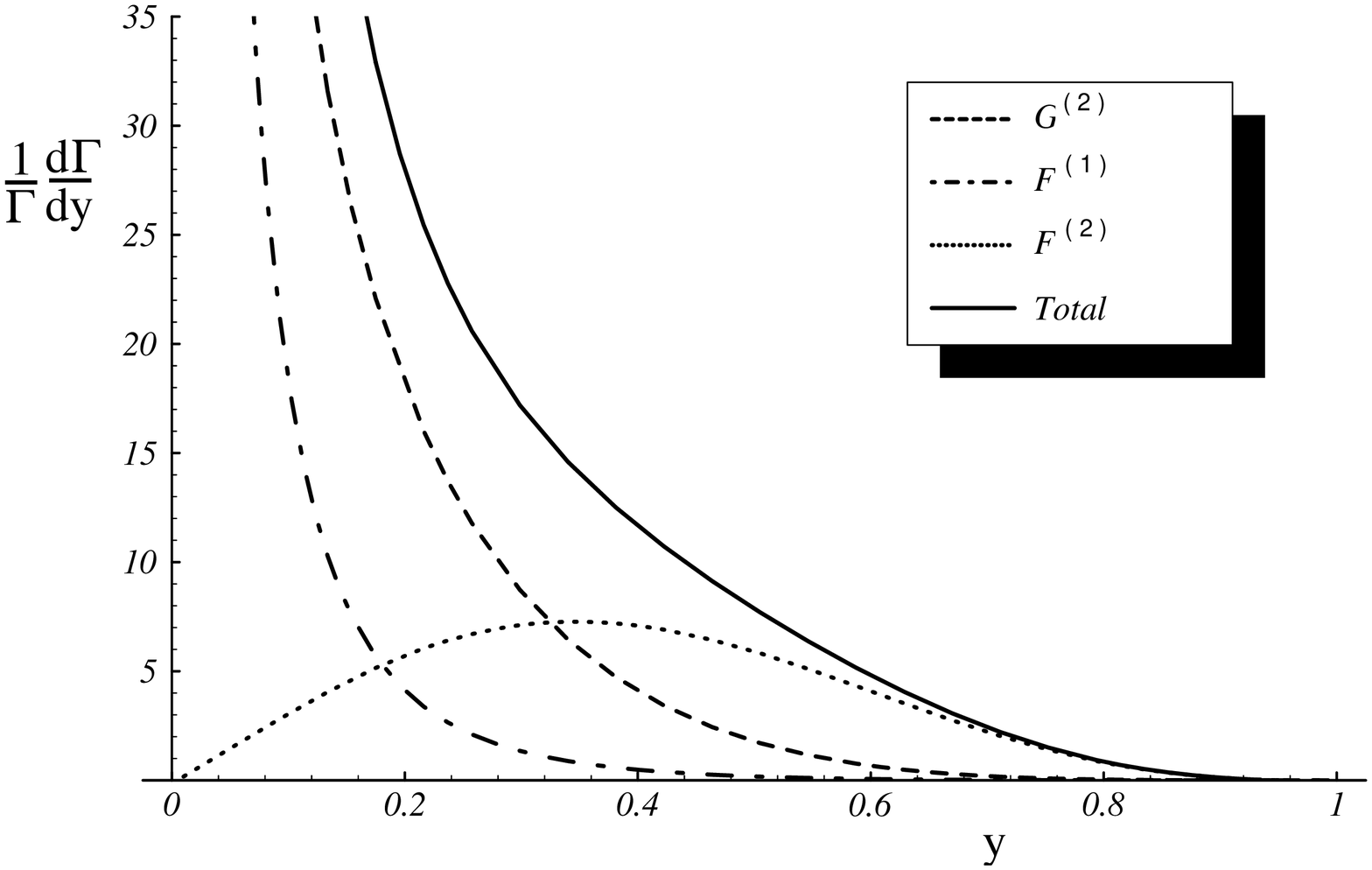}

\bye